\RequirePackage{fix-cm}
\documentclass[twocolumn]{svjour3}          

\usepackage{graphicx}

\usepackage{doi}
\usepackage{mwe}
\usepackage{lipsum}
\usepackage{subfig}
\usepackage[]{natbib}
\usepackage{epstopdf}
\usepackage{amsmath}
\usepackage{amsfonts}
\usepackage{enumitem}
\usepackage{color}
\usepackage{breakcites}
\usepackage{pgfplots}                           
\usepackage{pgfplotstable}
\usepackage{tikz}
\usepackage{rotating}

\definecolor{NewRed}{HTML}{990000}
\definecolor{NewBlue}{HTML}{0022BB}
\definecolor{NewGreen}{HTML}{00612C}
\definecolor{NewSand}{HTML}{EDEDCE}
\definecolor{NewTeal}{HTML}{1F5940}
\definecolor{NewOrange}{HTML}{C43B09}

\usepackage{etoolbox}
\makeatletter
\patchcmd\@combinedblfloats{\box\@outputbox}{\unvbox\@outputbox}{}{\errmessage{\noexpand patch failed}}
\makeatother

\begin{document}
\title{Simple single-scale microstructures based on optimal rank-3 laminates}

\author{E. Tr\"aff \and
O. Sigmund \and
J. P. Groen}


\institute{
E. Tr\"aff\and
O. Sigmund\and
J. P. Groen
\at
              Department of Mechanical Engineering, Solid Mechanics,
Technical University of Denmark, 2800 Kgs, Lyngby, Denmark \\
              Tel.: +45-45254252\\
              \email{jergro@mek.dtu.dk}           
}

\date{Received: date / Accepted: date}

\maketitle

\begin{abstract}
\begin{sloppypar}
With the goal of identifying optimal elastic single-scale microstructures for multiple loading situations, the paper shows that qualified starting guesses, based on knowledge of optimal rank-3 laminates, significantly improves chances of convergence to near optimal designs. Rank-3 laminates, optimal for a given set of anisotropic loading conditions, are approximated on a single scale using a simple mapping approach. We demonstrate that these mapped microstructures perform relatively close to theoretical energy bounds. Microstructures with performance even closer to the bounds can be obtained by using the approximated rank-3 structures in a further step as starting guesses for inverse homogenization problems. Due to the non-convex nature of inverse homogenization problems, the starting guesses based on rank-3 laminates outperform classical starting guesses with homogeneous or random material distributions. Furthermore, the obtained single-scale microstructures are relatively simple, which enhances manufacturability. Results, obtained for a wide range of loading cases, indicate that microstructures with performance within 5-8\% of the theoretical optima can be guarantied, as long as feature sizes are not limited by minimium size constraints.
\end{sloppypar}
\end{abstract} 

\section{Introduction}
For many engineering applications it is of interest to design periodic materials with tailored or extremal properties. The use of topology optimization to such design problems was introduced by~\citet{Bib:SigmundInverse1994} and is generally referred to as inverse homogenization. Since its introduction, the approach has been applied to many design problems such as materials with negative Poisson's ratio~\citep{Bib:LarsenSigmund97,Bib:AndreassenLazarovSigmund2014,Bib:ClausenPoison2015}, materials with maximum shear and bulk moduli~\citep{Bib:SigmundBone1999,Bib:SigmundExtremal2000}, or materials with increased buckling strength~\citep{Bib:Neves2002,Bib:ThomsenWangSigmund}.
Besides elasticity problems, the method has been successfully applied to design materials for $e.g.$ thermal, fluid, and  wave guiding problems, as well as a large number of other applications. A detailed discussion and overview of the field of material design is given in the recent review paper by~\citet{Bib:GuestReviewMatDesign}.

\begin{sloppypar}
In the context of elasticity, several researchers have looked into multi-scale or so-called hierarchical designs. Here, the topology optimization problem is divided into a global material distribution problem and local composite material design problems at the microscale. The composite microstructures are tailored for maximum strain energy, subject to the local stresses or strains~\citep{Bib:RodriguesHierarchical,Bib:CoelhoParallelHierarchical,Bib:XiaBreitkopf}. 
Hence, at each local point an inverse homogenization problem is solved to optimize the microstructures for one or more loading cases. To reduce the computational cost associated with the large number of inverse homogenization problems or to ensure less complex manufacturing, one can restrict the amount of unique microstructures; however, this comes with a loss of optimality~\citep{Bib:Liu2008,Bib:SchuryStingl2012,Bib:Sivapuram}. Anyway, when assuming separation of scales, multiscale design approaches require repeated and point-wise optimization of local microstructures, subjected to one or more load cases.

It is well-known that composite microstructures, assembled from two isotropic materials (possibly one of them being void), satisfying the optimal bound on complementary energy ($i.e.$ maximal strain energy) can be realized as so-called rank-$3$ laminates~\citep{Bib:LurieCherkaev84,Bib:FrancfortMurat86,Bib:Milton86,Bib:Avellaneda87}, for plane problems under the assumption of linear elasticity. For extensive details on rank-$N$ laminates and homogenization-based topology optimization the reader is referred the monographs by~\citet{Bib:AllaireBook,Bib:CherkaevBook,Bib:TopOptBook}. In fact,~\citet{Bib:Guedes2003} compared the performance of microstructures optimized using inverse homogenization with the energy bounds that can be reached using optimal rank-3 laminates. Interestingly, however, this background has never lead to a detailed study that aims at using knowledge of the optimal rank-3 laminates for the design of microstructures using inverse homogenization.

\begin{figure}[h!]
\centering
\includegraphics[width=0.5\textwidth]{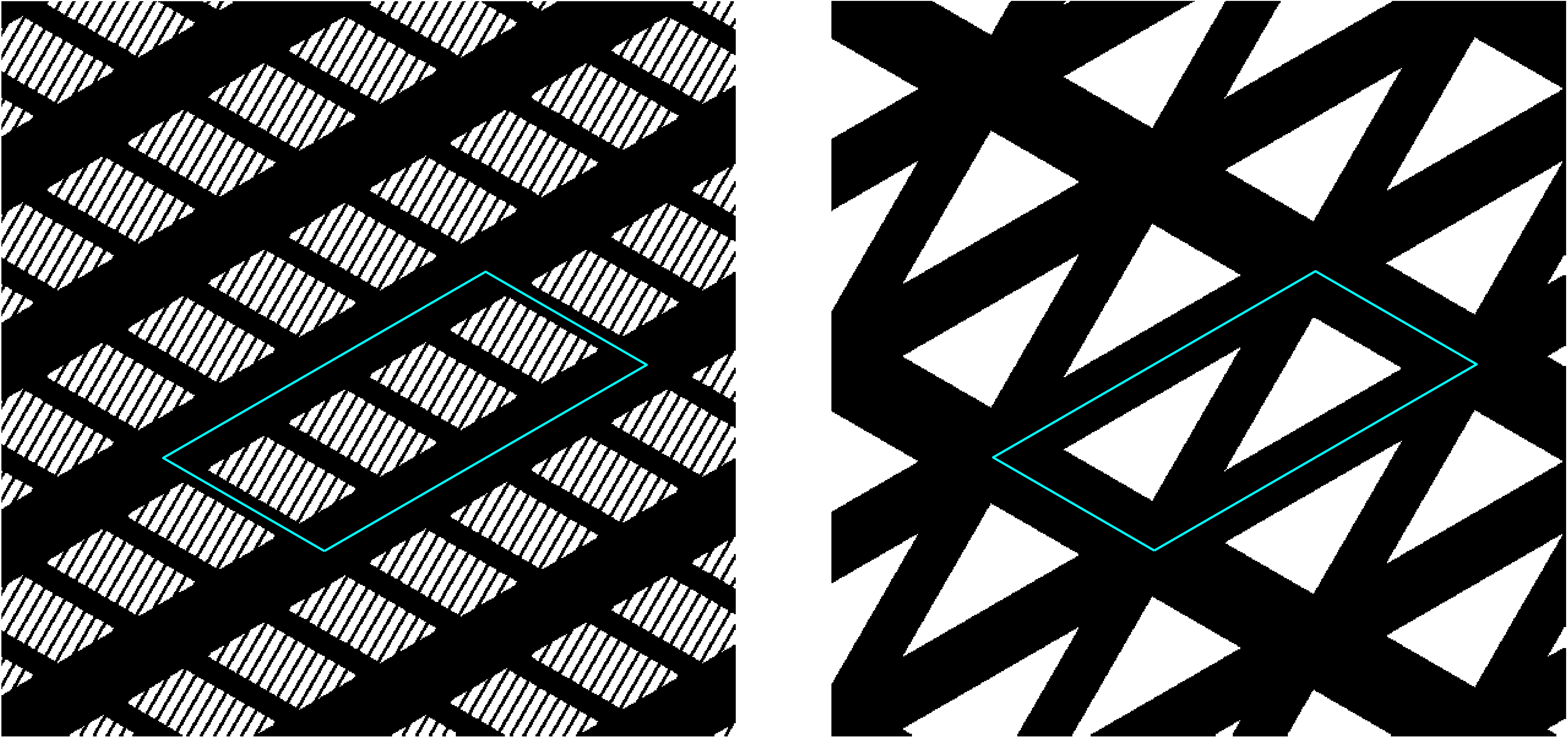}
\caption{Left: rank-3 microstructure with indicated hierarchy, Right: approximated single-scale microstructure. Illustrations are based on a rank-3 laminate with, $\theta_{1} = \pi/3$, $\mu_{1} = 0.2$, $\theta_{2} = -\pi/6$, $\mu_{2} = 0.25$, $\theta_{3} = 1\pi/6$ and $\mu_{3} = 0.50$.}
\label{Fig:Intro.1}
\end{figure}

Apart from rank-$N$ laminates, single-scale, so-called Vigdergauz microstructures have been conceived and analyzed by analytical means \citep{Vig94-01,Vig94,GraKoh95,LiuJamLeo07}, or have come out as result of many topology optimization approaches, $e.g.$~\citep{Bib:SigmundBone1999,Bib:SigmundExtremal2000}. Vigdergauz structures may achieve the maximum bulk modulus bound ($i.e.$ are optimal for hydrostatic loading), but cannot achieve the maximum shear modulus bound simultaneously ($i.e.$ a single scale structure cannot produce the theoretically maximal isotropic Young's modulus). A proof that more length-scales are required to reach the theoretical bounds is given by~\citet{Bib:AllaireAubry1999}.  Hence, it is clear that imposing a single-scale constraint on microstructure design inherently will put a limit on achievable performance, but on the other hand, this optimality gap may be acceptable considering that the microstructures become manufacturable. For other works and discussions on single-scale microstructures with respect to theoretical bounds, the readers are referred to \citep{BouKoh08,SigAagAnd16,BerWadMcM17}.
\end{sloppypar}
In this work, we propose a systematic approach for creating near-optimal single-scale microstructures, based on the layer directions and relative widths of optimal rank-3 laminates, see $e.g.$ Figure~\ref{Fig:Intro.1}. We demonstrate that this approximation of the microstructures already performs relatively close, $e.g.$ 5-15$\%$, to the optimal energy bounds. Furthermore, we demonstrate that this performance can be further improved by using these single-scale microstructures as starting guess for an inverse homogenization problem. In general, the non-convex nature of the inverse homogenization problem results in convergence to inferior or very complex microstructures when using traditional starting guesses, such as uniform or random material distributions. It is shown that the proposed starting guesses from rank-3 laminates perform overall better than conventional starting guesses, especially when lower volume fractions are considered. Besides this better performance, the optimized microstructures using rank-3 inspired starting guesses are geometrically much simpler, in turn facilitating a simpler manufacturing process.


\section{Interpreting optimal microstructures on a single scale}
\label{Sec:MomentForm}
\begin{sloppypar}
The theory to obtain optimal rank-$N$ laminates is well-established, as mentioned before, and only included here for completeness. In this work we parameterize the optimal rank-3 laminate in terms of $4$ trigonometric moments as introduced by~\citet{Bib:MiltonAvellaneda89}. Afterwards, we use the method of~\citet{Bib:Lipton1994MomentRecon} to reconstruct a rank-3 laminate from the optimal moments. Finally, we propose a novel method to approximate the rank-3 microstructure by a single-scale microstructure.
\end{sloppypar}

\subsection{Optimal microstructures by the moment formulation}
We seek to minimize the complementary work $\mathcal{C}$ on a periodic microstructure subjected to $n_{s}$ stress cases.
\begin{equation} \label{Eq:Moment.1}
\mathcal{C} = \frac{1}{2}\sum_{j}^{n_{s}} w_{j} \boldsymbol{\sigma}_{j} :\boldsymbol{C}^{H}: \boldsymbol{\sigma}_{j},
\end{equation}
where, $\boldsymbol{\sigma}_{j}$ is the stress tensor corresponding to the $j$-$th$ load case, $w_{j}$ is the relative weighting such that $\sum_{j=1}^{n_{s}} w_{j} = 1$, and $\boldsymbol{C}^{H}$ describes the effective material compliance tensor of the considered microstructure. For a finite-rank microstructure the effective compliance tensor can be written as,
\begin{equation} \label{Eq:Moment.2}
\begin{aligned}
& \boldsymbol{C}^{H} = \boldsymbol{C}^{+} - (1-f)\\
& \Big((\boldsymbol{C}^{+}-\boldsymbol{C}^{-})^{-1} -f E^{+} \sum_{n=1}^{N} p_{n} (\boldsymbol{t}_{n}\otimes\boldsymbol{t}_{n})\otimes (\boldsymbol{t}_{n}\otimes\boldsymbol{t}_{n})
\Big)^{-1},
\end{aligned}
\end{equation}
with $\boldsymbol{C}^{+}$ and $\boldsymbol{C}^{-}$ being the properties of the stiff (+) and compliant (-) material respectively, $f$ is the volume fraction of the stiff material, while $E^{+}$ is its corresponding Young's modulus. Furthermore, $\boldsymbol{t}_{n}$ represents the tangent vector of layer $n$, while $p_{n}\geq0$ describes the relative contribution of the $n$-$th$ layer, s.t. $\sum_{n=1}^{N} p_{n} = 1$. Finally, $\otimes$ indicates the dyadic product.

For convenience we can rewrite the 4-$th$ rank tensor using the following orthogonal basis of second-rank tensors,
\begin{equation} \label{Eq:Moment.3}
    \begin{aligned}
        \boldsymbol{\xi}_{1} = \frac{1}{\sqrt{2}} \begin{bmatrix} 1 & 0 \\ 0 & -1  \end{bmatrix} &,&\boldsymbol{\xi}_{2} = \frac{1}{\sqrt{2}} \begin{bmatrix} 0 & 1 \\ 1 & 0  \end{bmatrix} &,&\boldsymbol{\xi}_{3} = \frac{1}{\sqrt{2}} \begin{bmatrix} 1 & 0 \\ 0 & 1  \end{bmatrix}.
    \end{aligned}
\end{equation}
Furthermore, we can reduce the number of variables describing the effective laminate properties using the following four moments,
\begin{equation} \label{Eq:Moment.4}
    \begin{aligned}
    m_{1} = \sum_{n=1}^{N}p_{n} \text{cos}(2\theta_{n}), && m_{2} = \sum_{n=1}^{N}p_{n} \text{sin}(2\theta_{n}), \\
    m_{3} = \sum_{n=1}^{N}p_{n} \text{cos}(4\theta_{n}), && m_{4} = \sum_{n=1}^{N}p_{n} \text{sin}(4\theta_{n}). \\
	\end{aligned}
\end{equation}
Hence, we can write
\begin{equation} \label{Eq:Moment.5}
\begin{aligned}
&\boldsymbol{\xi} :\big(\sum_{n=1}^{N} p_{n} (\boldsymbol{t}_{n}\otimes\boldsymbol{t}_{n})\otimes (\boldsymbol{t}_{n}\otimes\boldsymbol{t}_{n})\big): \boldsymbol{\xi}= \boldsymbol{M},  \\ &\boldsymbol{M}  = \frac{1}{4}\begin{bmatrix}
1+m_{3} & m_{4} & -2m_{1} \\
& 1-m_{3} & -2m_{2} \\
& & 2
\end{bmatrix}.
\end{aligned}
\end{equation}
As described by \citet{Bib:KreinNudelman77}, the feasible set of moments $\mathcal{M}$ can be described as,
\begin{equation} \label{Eq:Moment.6}
\begin{aligned}
\mathcal{M} &= \boldsymbol{m} \in \mathbb{R}^{4},   \\
&s.t. \begin{cases}
m_{1}^{2} + m_{2}^{2} \leq 1,  \\
-1 \leq m_{3} \leq 1, \\
\frac{2m_{1}^{2}}{1+m_{3}}+\frac{2m_{2}^{2}}{1-m_{3}}+ \frac{m_{4}^{2}}{1-m_{3}^{2}}  - \frac{4m_{1}m_{2}m_{4}}{1-m_{3}^{2}} \leq 1.
\end{cases}
\end{aligned}
\end{equation}
Hence, for a given set of loadings conditions, and material properties $f$, $E^{+}$ and $E^{-} = 10^{-9} E^{+}$, we can find the optimal set of moments, minimizing $\mathcal{C}$, to find the optimal effective compliance tensor. As discussed by~\citet{Bib:Lipton1994MomentRecon}, this is a convex problem with respect to the moments. Furthermore, \citet{Bib:Lipton1994MomentRecon} proposed a procedure to reconstruct the relative layer widths $\mu_{n}$ and orientations $\theta_{n}$ of a rank-3 laminate from the optimal moments, which is for convenience repeated in Appendix~\ref{Sec:Appendix.recon}

\subsection{Approximation of a rank-3 laminate on a single scale}
We abandon the separation of length-scales specific to rank-3 laminates and try to approximate the 3-layered composite on a single scale. To do so we need: 1)  to create a periodic unit-cell respecting the layer orientations, 2) to make sure that the relative contributions of the different layer widths are preserved on the single scale.

\begin{figure}[h!]
\centering
\includegraphics[width=0.5\textwidth]{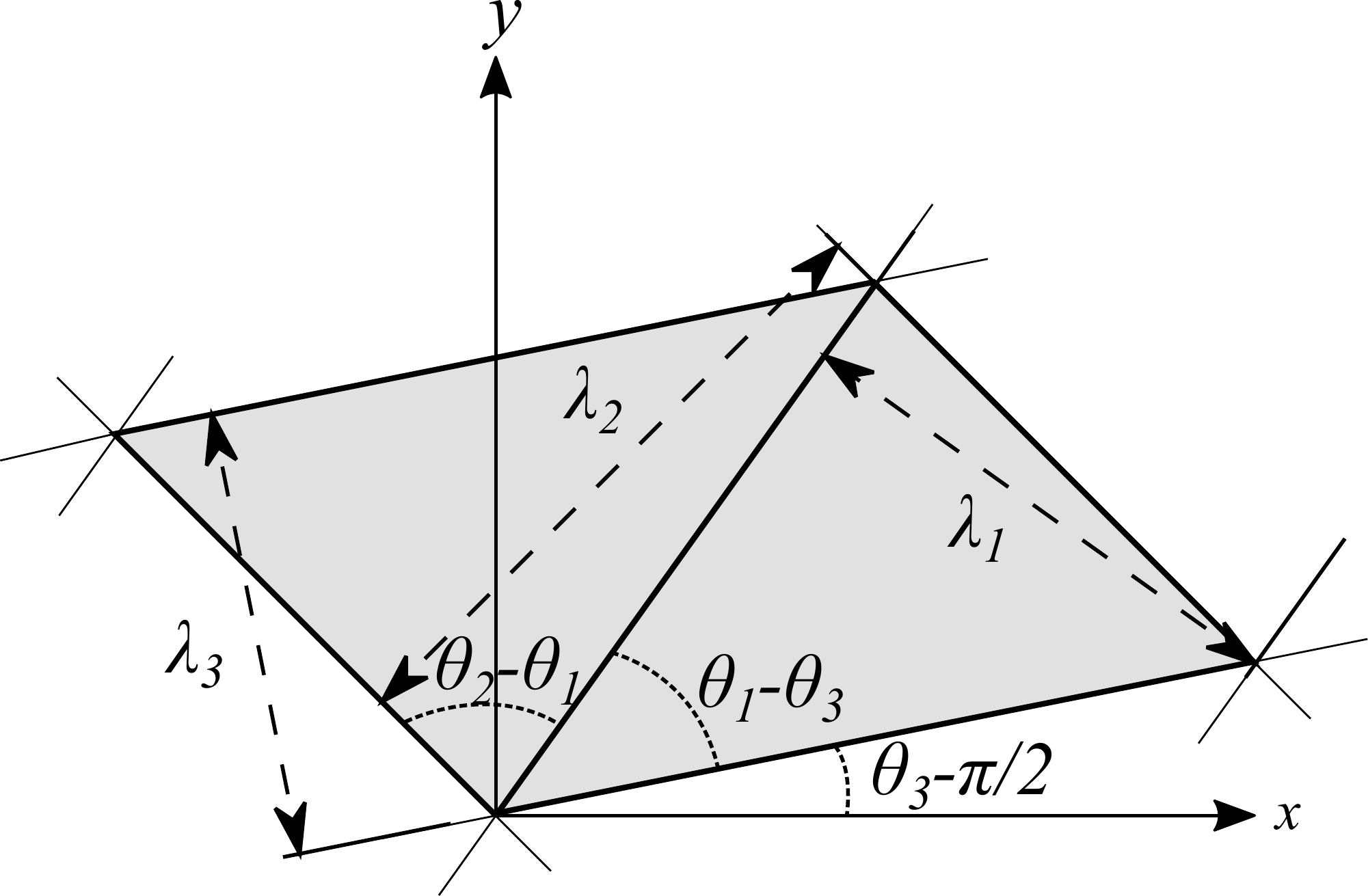}
\caption{Parallelogram used as unit-cell and corresponding dimensions.}
\label{Fig:Approx.1}
\end{figure}

To transform a rank-3 layered composite into a single-scale periodic lattice structure (see illustration in Figure~\ref{Fig:Intro.1}), the spacing of each layer $\lambda_{n}$ has to be adapted, such that the unit-cell can be described using a parallelogram, as shown in Figure~\ref{Fig:Approx.1}.

In this work layer 3 is set to be the base layer, with a layer spacing $\lambda_{3}=1$. Using standard geometric relations, the other two layer spacings can be calculated,
\begin{equation}\label{Eq:Approx.1}
\begin{aligned}
\lambda_{1} = \left|\text{sin}(\theta_{1}-\theta_{3})\right| \left|\frac{\lambda_{3}}{\text{tan}(\theta_{2}-\theta_{3})}-\frac{\lambda_{3}}{\text{tan}(\theta_{1}-\theta_{3})} \right|, \\
\lambda_{2} =  \left|\text{sin}(\theta_{2}-\theta_{3})\right| \left|\frac{\lambda_{3}}{\text{tan}(\theta_{2}-\theta_{3})}-\frac{\lambda_{3}}{\text{tan}(\theta_{1}-\theta_{3})} \right|.
\end{aligned}
\end{equation}
The area $A$ of the parallelogram is described as,
\begin{equation}\label{Eq:Approx.2}
A =  \left|\frac{\lambda_{3}^{2}}{\text{tan}(\theta_{2}-\theta_{3})}-\frac{\lambda_{3}^{2}}{\text{tan}(\theta_{1}-\theta_{3})} \right|.
\end{equation}
To make the unit-cell have a unit-area, the corresponding layer spacings are normalized to,
\begin{equation}\label{Eq:Approx.3}
\begin{aligned}
\tilde{\lambda}_{1} = \frac{\lambda_{1}}{\sqrt{A}}, && \tilde{\lambda}_{2} = \frac{\lambda_{2}}{\sqrt{A}}, && \tilde{\lambda}_{3} = \frac{\lambda_{3}}{\sqrt{A}}.
\end{aligned}
\end{equation}
For the case that the reconstructed laminate is only a rank-2 laminate, we set the corresponding layer spacings equal, while maintaining a unit-area.

For the layer widths, we cannot directly use the values $\mu_{1}$, $\mu_{2}$ and $\mu_{3}$, which are the relative layer widths of the stiff material (+) in each layer, since these values depend on the layer ordering. Therefore, we make use of the parameters that describe the relative contribution of each layer, $i.e.$ $p_{n}$ for $n=1,2,3$. The width of each layer used on the single scale $w_{n}$, are then obtained as,
\begin{equation} \label{Eq:Approx.4}
\begin{aligned}
w_{n} = \psi p_{n}, && n=1,2,3.
\end{aligned}
\end{equation}
Here $\psi$ is a scaling parameter found using a bi-section algorithm, such that the projected microstructure has the same volume fraction $f$ as the stiff material in the rank-3 laminate.

With the layer widths and relative spacings known, we can project the periodic microstructure. For this we use an approach similar to the recently proposed projection approaches presented in~~\citet{Bib:PantzTrabelsi,Bib:GroenSigmund2017,BiB:GDondersAllairePantz2018}. A periodic sequence of layer $n$ on a single scale has a material distribution $\tilde{\rho}_{n}$ which can be described as,
\begin{equation}\label{Eq:Approx.5}
\tilde{\rho}_{n}(\boldsymbol{x}) = H\Big(\text{cos}(\frac{2\pi(\boldsymbol{n}_{n} \cdot \boldsymbol{x})}{\tilde{\lambda}_{n}})- \text{cos}(\pi w_{n})
\Big).
\end{equation}
Here, $\boldsymbol{x}$ describes the spatial location, $\boldsymbol{n}_{n}$ describes the layer normal of layer $n$, and $H$ is the Heaviside function. Finally, the individual layer contributions can be combined in the total density distribution of the microstructure as,
\begin{equation}\label{Eq:Approx.6}
\rho(\boldsymbol{x}) = \text{min}\left\{\tilde{\rho}_{1}(\boldsymbol{x})+\tilde{\rho}_{2}(\boldsymbol{x})+\tilde{\rho}_{3}(\boldsymbol{x}),1 \right\}.
\end{equation}
Figure~\ref{Fig:Intro.1} shows an example of a resulting single-scale microstructure.

\section{Optimization for single-scale microstructures}
\label{Sec:Ihomog}
The unit-cell description given above can be used to parameterize the unit-cell using a discrete number of elements, such that density-based topology optimization can be performed. To tailor the microstructure properties such that the weighted complementary energy $\mathcal{C}$ is minimized an inverse homogenization approach is used. The topology optimization problem is solved in nested form, $i.e.$ for each design iteration and given density distribution, the homogenized constitutive properties, as well as the objective function and its corresponding derivatives are calculated. Afterwards, the material distribution is updated based on the gradients. The discretized optimization problem can thus be written as,
\begin{equation}\label{Eq:Ihomog.1}
\begin{aligned}
 & & \displaystyle \min_{\boldsymbol{\rho}} & :   \mathcal{C}(\boldsymbol{\rho}) = \frac{1}{2}\sum_{j}^{n_{s}} w_{j} \boldsymbol{\sigma}_{j} :\boldsymbol{C}^{H}(\boldsymbol{\hat{\bar{\rho}}}(\boldsymbol{\rho})): \boldsymbol{\sigma}_{j},     \\
 & & \textrm{s.t.}  		   & :   \boldsymbol{v}^{T} \boldsymbol{\hat{\bar{\rho}}}(\boldsymbol{\rho}) - f \leq 0,  	\\
 & &                           & :     \boldsymbol{0}  \leq  \boldsymbol{\rho}  \leq  \boldsymbol{1}.		\\
\end{aligned}
\end{equation}
\begin{sloppypar}
Here $\boldsymbol{v}$ is a vector containing the volumes of each density element, and $\boldsymbol{\hat{\bar{\rho}}}$ is a vector containing the physical densities. $\boldsymbol{C}^{H}(\boldsymbol{\hat{\bar{\rho}}}(\boldsymbol{\rho}))$ is the homogenized compliance tensor. To obtain this tensor we make use of the publicly available MATLAB code of~\citet{Bib:AndreassenAndreasenHomo}, which is modified to return the sensitivities of the objective w.r.t. design variables. More details regarding the sensitivity analysis for inverse homogenization problems can be found in~\citet{Bib:SigmundInverse1994,Bib:Guedes2003,Bib:TopOptBook} . Furthermore, we use the MATLAB implementation of the Method of Moving Asymptotes (MMA) introduced by~\citet{Bib:MMA}.

To ensure a well-posed problem, control of the minimum feature size of solid regions, as well as discreteness of optimized designs, the physical densities $\boldsymbol{\hat{\bar{\rho}}}$, are related to design variables $\boldsymbol{\rho}$ using the Heaviside projection scheme introduced by~\cite{Bib:GuestHeaviside}. The scheme is based on a filter radius $R$, which is slightly modified to take the periodicity of the unit-cell into account. A standard continuation approach is used for the projection parameter $\beta$ to improve convergence.
\end{sloppypar} 

\section{Numerical examples}
\label{Sec:Example}
In all examples, three different starting guesses are used for the inverse homogenization problems. First of all, the rank-3 laminate mapped on a single scale is used (\textit{Mapped SG}). The performance of the optimized design is compared to a design obtained with a starting guess containing a random density field (\textit{Random SG}), as well as a starting guess using a homogeneous density field (\textit{Homog. SG}). Furthermore, we directly evaluate the performance of the rank-3 microstructure approximated on the single scale (\textit{Mapped Rank-3}), and we compare all results to the energy bound given by the optimal rank-3 laminate (\textit{Rank-3}).

The \textit{Random SG} is obtained by generating random numbers between $0$ and $\min(1, 2f)$. A density filter is applied to the densities to ensure some continuity in the initial guess. The \textit{Homog. SG} is not truly homogeneous, otherwise numeric problems will arise from having identical sensitivity values for all elements. Instead, all elements are set to the value of $f$, except for the elements within a filter radius $R$ of the unit-cell center, which are set to {0}. In this manner the starting guess is perturbed independently of mesh refinement.

Both the \textit{Random SG}, and the \textit{Homog. SG} are optimized in a unit-cell of dimensions $1$ by $1$. To make comparisons with the rank-3 inspired parallelogram unit-cell starting guesses more fair, we also tried to include the angle between the axes as an additional design variable. However, this increased design freedom did in general not lead to better designs; hence, we stick with a square unit-cell for the \textit{Random SG} and \textit{Homog. SG} in the following.

\subsection{Example 1: Small filter radius}
\begin{figure}
\centering
\includegraphics[width=0.8\linewidth]{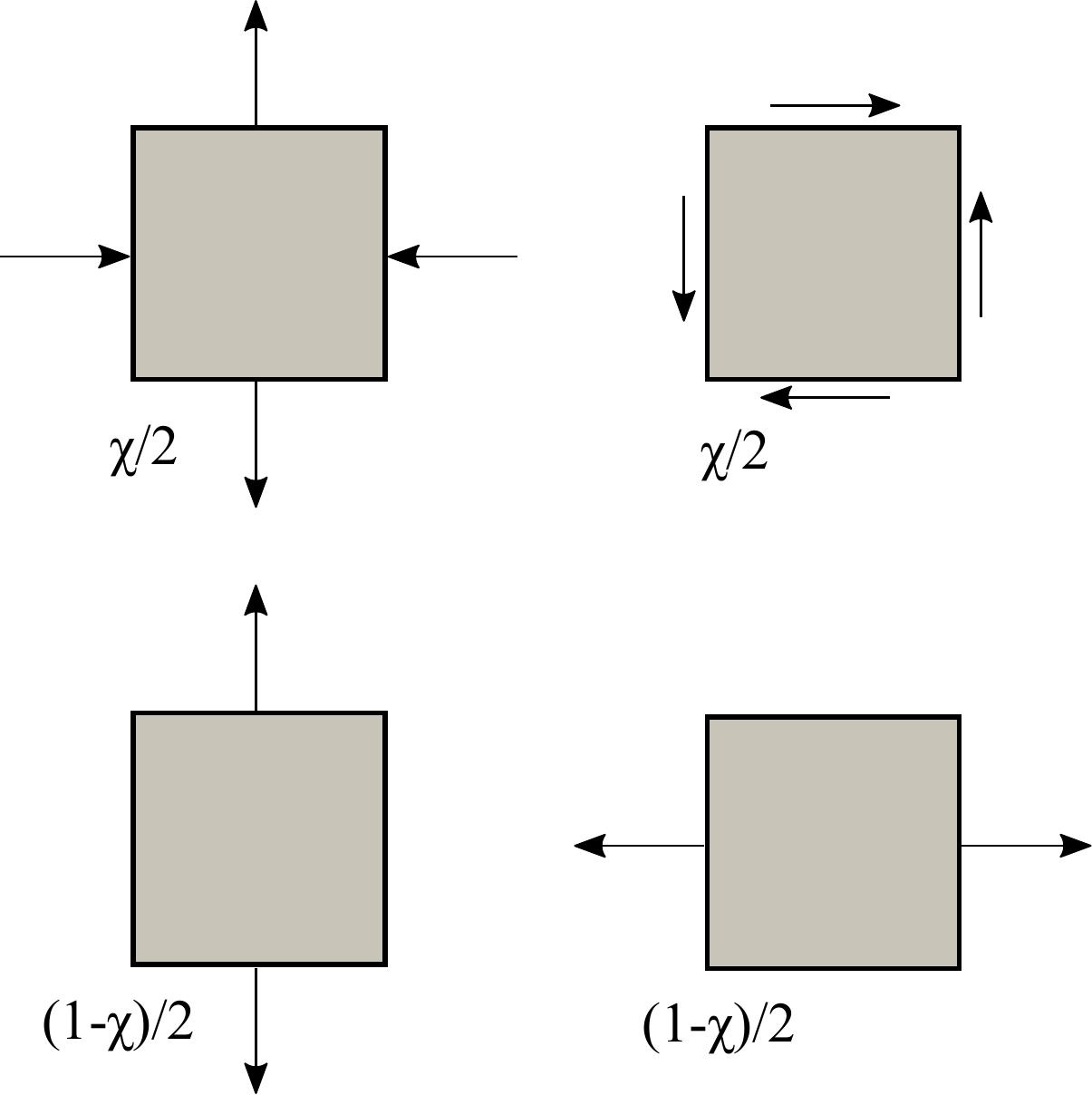}
\caption{Illustration of the four stress cases and their respective weights used in examples 1,2 and 3.}
\label{Fig:Example.1}
\end{figure}

This first three examples are based on four stress cases, which are weighted using parameter $\chi \in [0,1]$ as proposed by \citet{Bib:Guedes2003}. The four cases, with their respective weight are shown in Figure~\ref{Fig:Example.1}, and can be written out as,
\begin{equation} \label{Eq:Example.0}
\begin{aligned}
w_1 \sigma_1 &= \frac{\chi}{2} \begin{bmatrix} -1 & 0\\ 0 & 1\end{bmatrix}, &&&
w_2 \sigma_2 &= \frac{\chi}{2} \begin{bmatrix} 0 &1 \\1& 0\end{bmatrix}, \\
w_3 \sigma_3 &= \frac{1 - \chi}{2} \begin{bmatrix} 1 & 0 \\ 0 & 0\end{bmatrix}, &&&
w_4 \sigma_4 &= \frac{1 - \chi}{2} \begin{bmatrix} 0& 0\\ 0 & 1\end{bmatrix}.
\end{aligned}
\end{equation}
The enforced length-scale in this example is $2R=0.05$, corresponding to $5\%$ of the side length in a square unit-cell. This small feature size allows for quite fine structural members. Furthermore, a volume fraction of $f=0.5$ is used, and the unit-cells are discretized using $200\times200$ elements.

\begin{figure}
\centering
\begin{tikzpicture}[scale=1]
\footnotesize
\begin{axis}[
    		    grid=major,
                width=\linewidth, 
                height=6cm,
                scaled x ticks = false,
                /pgf/number format/.cd,
                set thousands separator={ },
                xlabel = {$\chi$},
                ylabel = {Objective values $\mathcal{C}$},
                xmin=-0.01,xmax=1.01,
                legend entries={
                    Mapped SG,
                    Random SG,
                    Homog. SG,
                    Rank-3,
                    Mapped Rank-3
                },
                legend style={
                    draw=none,
                    legend columns=3,
                    at={(1.05,1.02)},
                    anchor=south east,
                    /tikz/every even column/.append style={column sep=0.5cm}
                },
                ytick = {0,2.5,5,7.5,10,12.5},
                xtick = {0,.1,.2,.3,0.4,.5,0.6,.7,0.8,.9,1}
    ] 
        
    
    \addplot[mark=x,mark size=3,NewRed,only marks,thick] table[x index=0, y index=3] {ex1_2.txt};
    
    \addplot[mark=triangle,mark size=3,NewBlue,only marks,thick] table[x index=0, y index=4] {ex1_2.txt};
    
    \addplot[mark=o,mark size=2,NewTeal,only marks,thick] table[x index=0, y index=5] {ex1_2.txt};
    
    \addplot[sharp plot,black,thick] table[x index=0, y index=1] {ex1_2.txt};
    
    \addplot[sharp plot,dashed,black] table[x index=0, y index=2] {ex1_2.txt};

\end{axis}
\end{tikzpicture}
\begin{tikzpicture}[scale=1]
\footnotesize
\begin{axis}[
    		    grid=major,
                width=\linewidth, 
                height=6cm,
                scaled x ticks = false,
                /pgf/number format/.cd,
                set thousands separator={ },
                xlabel = {$\chi$},
                ylabel = {Scaled objective values},
                xmin=-0.01,xmax=1.01,
                legend style={
                    draw=none,
                    legend columns=3,
                    at={(1.05,1.02)},
                    anchor=south east,
                    /tikz/every even column/.append style={column sep=0.5cm}
                },
                xtick = {0,.1,.2,.3,0.4,.5,0.6,.7,0.8,.9,1}
    ] 
        
    
    \addplot[mark=x,mark size=3,NewRed,only marks,thick] table[x index=0, y index=3] {ex1.txt};
    
    \addplot[mark=triangle,mark size=3,NewBlue,only marks,thick] table[x index=0, y index=4] {ex1.txt};
    
    \addplot[mark=o,mark size=2,NewTeal,only marks,thick] table[x index=0, y index=5] {ex1.txt};
    
    \addplot[sharp plot,black,thick] table[x index=0, y index=1] {ex1.txt};
    
    \addplot[sharp plot,dashed,black] table[x index=0, y index=2] {ex1.txt};

\end{axis}
\end{tikzpicture}
\caption{Top: Resulting complimentary energy $\mathcal{C}$ for different values of $\chi$, $f=0.5$, and a length-scale of $0.05$. Bottom: Values normalized using the rank-3 energy bound.}
\label{Fig:Example.2}
\end{figure}
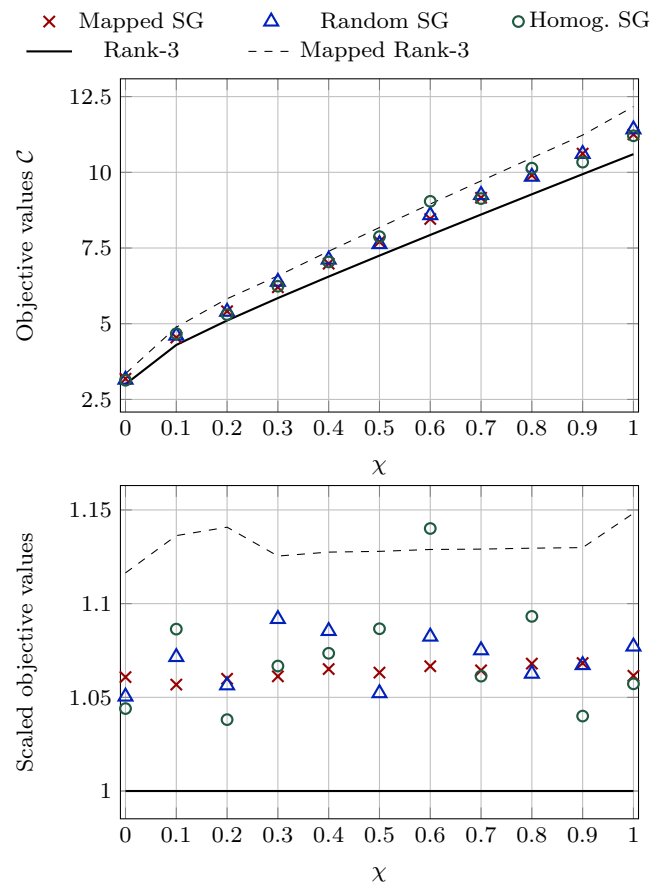

The found objective values $\mathcal{C}$ for different starting guesses and $11$ equally spaced values of $\chi$ are shown in the top of Figure~\ref{Fig:Example.2}. These results are in agreement with the results obtained by \citet{Bib:Guedes2003}. In the normalized plot shown in the bottom of Figure~\ref{Fig:Example.2}, starting guesses are compared based on their relative performance, $i.e.$ the results are normalized using the energy bound of the rank-3 microstructure.

\begin{figure}[h!]
\centering
\subfloat[Mapped Rank-3.]{\includegraphics[width=0.45\linewidth]{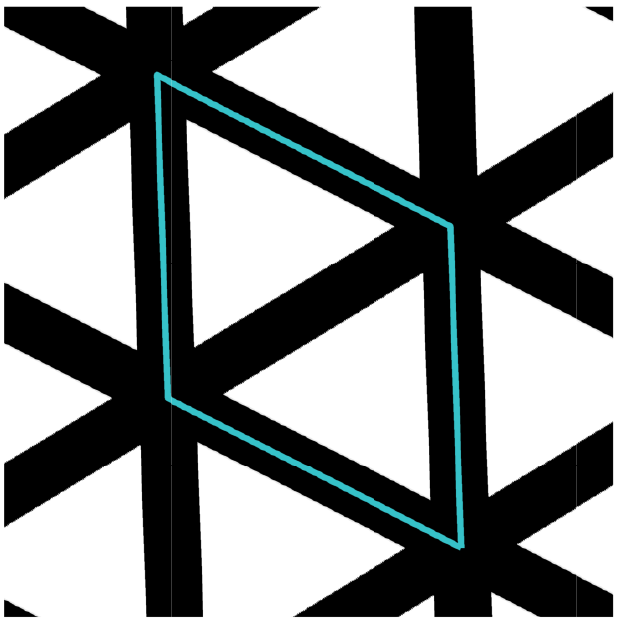}}\label{Fig:Example.3.1} \quad
\subfloat[Mapped SG.]{\includegraphics[width=0.45\linewidth]{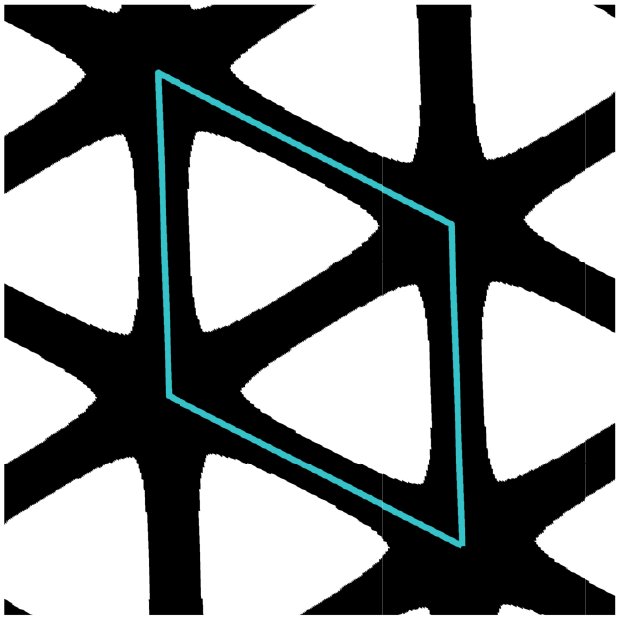}}\label{Fig:Example.3.2} \\
\subfloat[Random SG.]{\includegraphics[width=0.45\linewidth]{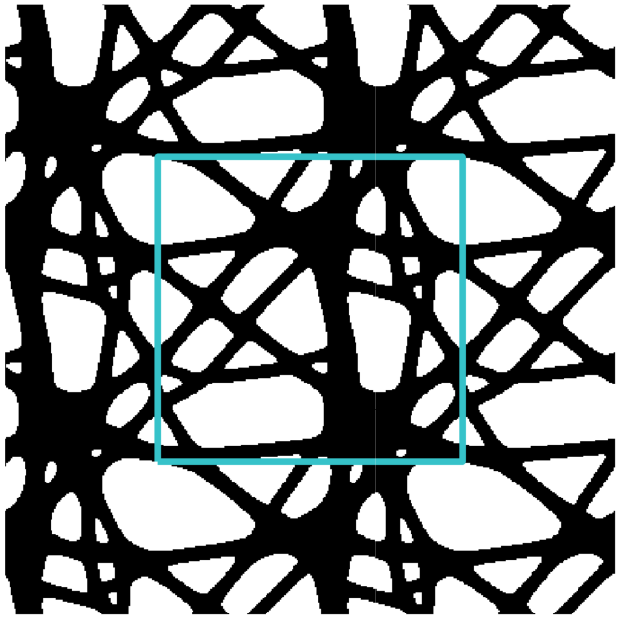}}\label{Fig:Example.3.3}  \quad
\subfloat[Homog. SG.]{\includegraphics[width=0.45\linewidth]{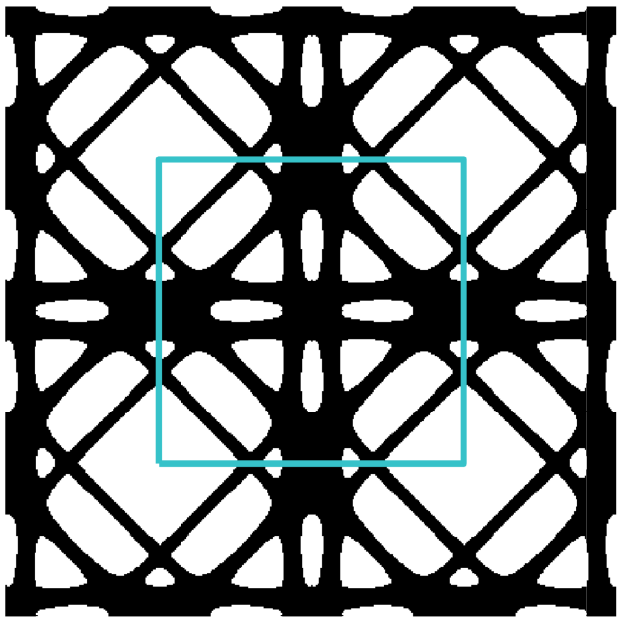}}\label{Fig:Example.3.4} \\
\caption{Resulting structures shown in a domain of size $2 \times 2$ with the unit-cell highlighted, for $\chi=0.7$, $f=0.5$, and a length-scale of $0.05$.}
\label{Fig:Example.3}
\end{figure}

From the objective values observed in this example it is not yet clear that the proposed starting guess has much benefit. Only when $\chi = 0.1$, $\chi = 0.3$, $\chi = 0.4$ and $\chi = 0.6$ the $\textit{Mapped SG}$ outperforms the other two starting guesses. The single-scale microstructures corresponding to  $\chi=0.7$ are shown in Figure~\ref{Fig:Example.3}. It is clear from Figure~\ref{Fig:Example.3}(c) that using \textit{Random SG} results in somewhat chaotic looking optimized microstructures, which, although well-performing, may become problematic in terms of manufacturability. Therefore, we focus in the remainder on a comparison between the use of \textit{Mapped SG} and \textit{Homog. SG}

Furthermore, it can be observed that the structures generated with \textit{Homog. SG} are more complex and consist of many members at the smallest allowable feature size. On the other hand, the microstructures generated with \textit{Mapped SG} have few large design features, while being only marginally worse in terms of performance. These fewer and larger design features may be desirable from a manufacturing viewpoint.

\begin{figure}[h!]
\centering
\subfloat[Mapped SG.]{\includegraphics[width=0.45\linewidth]{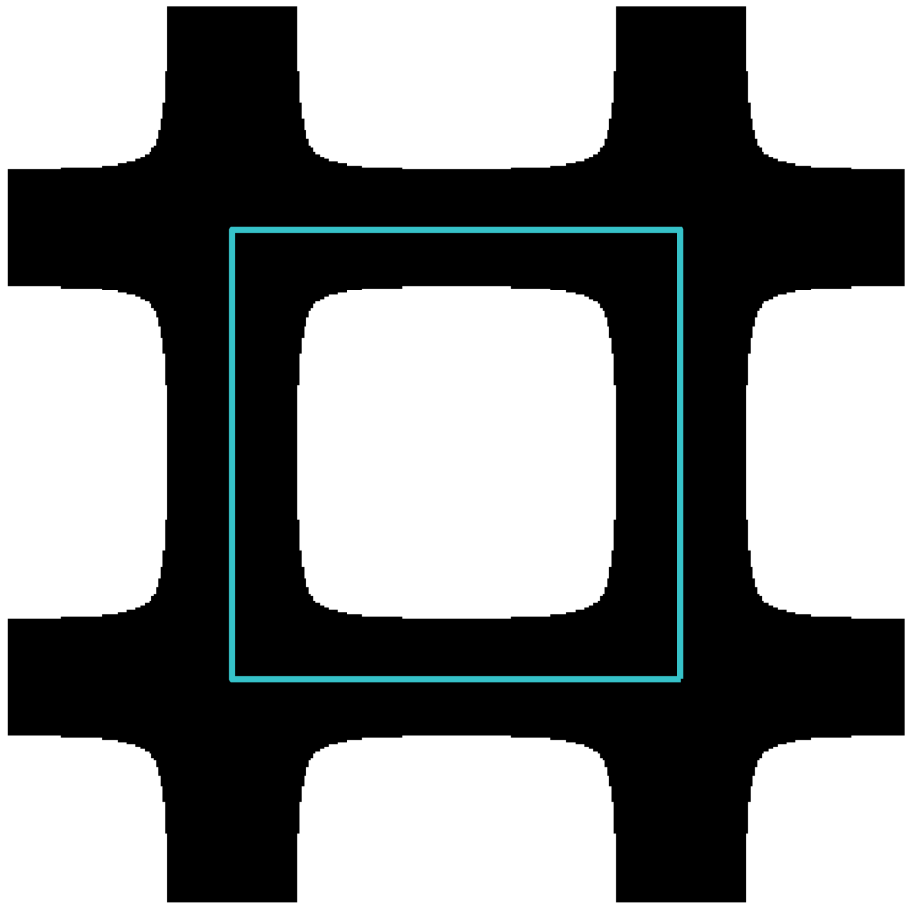}} \quad
\subfloat[Homog. SG.]{\includegraphics[width=0.45\linewidth]{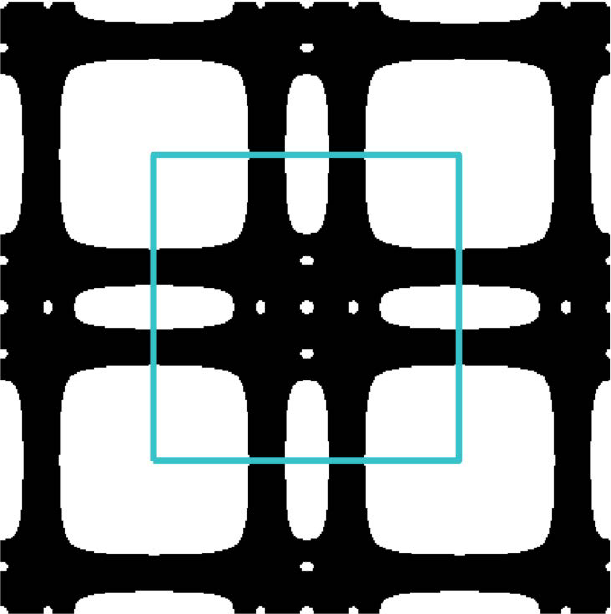}} \\
\subfloat[Improved mapping.]{\includegraphics[width=0.45\linewidth]{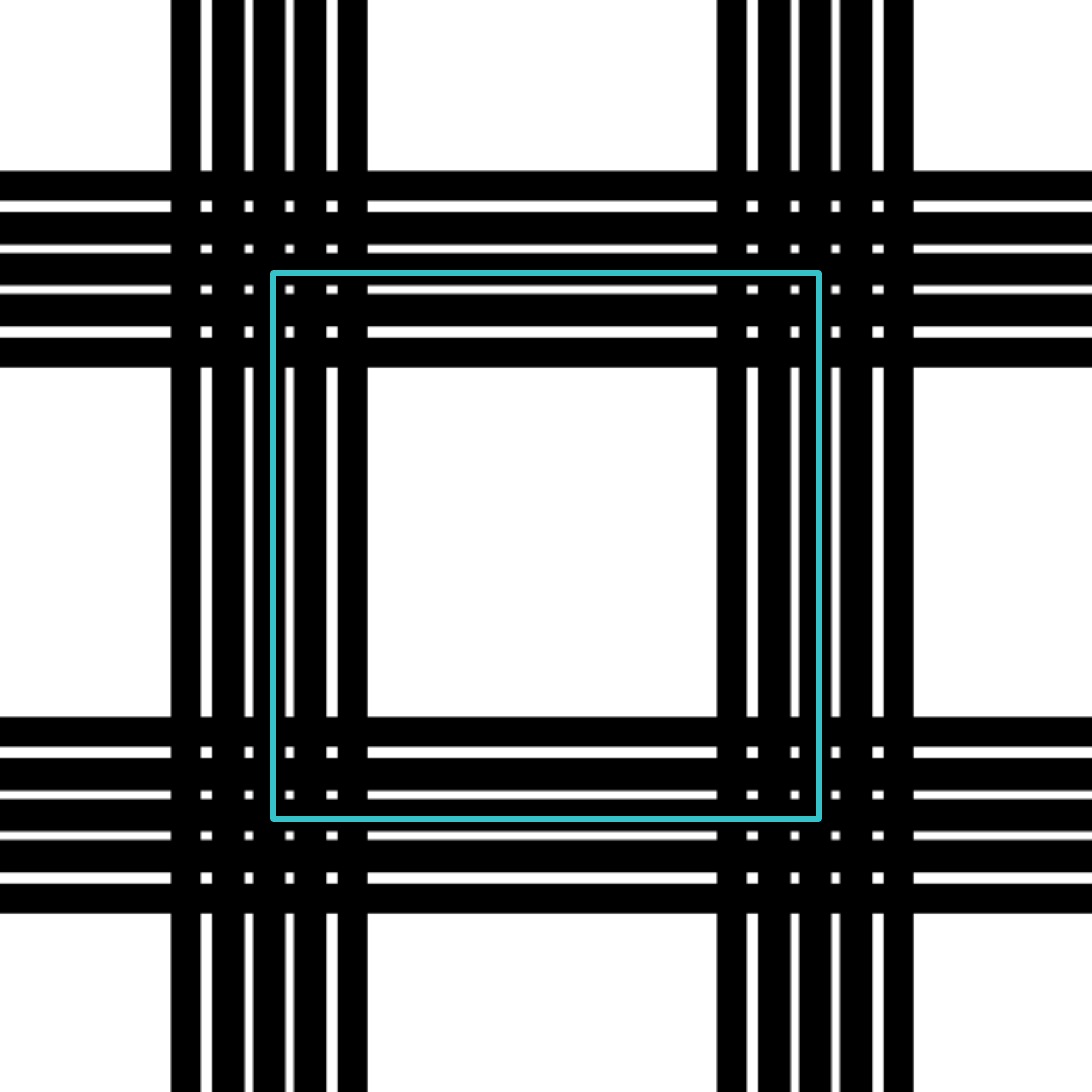}} \quad
\subfloat[Improved mapped SG.]{\includegraphics[width=0.45\linewidth]{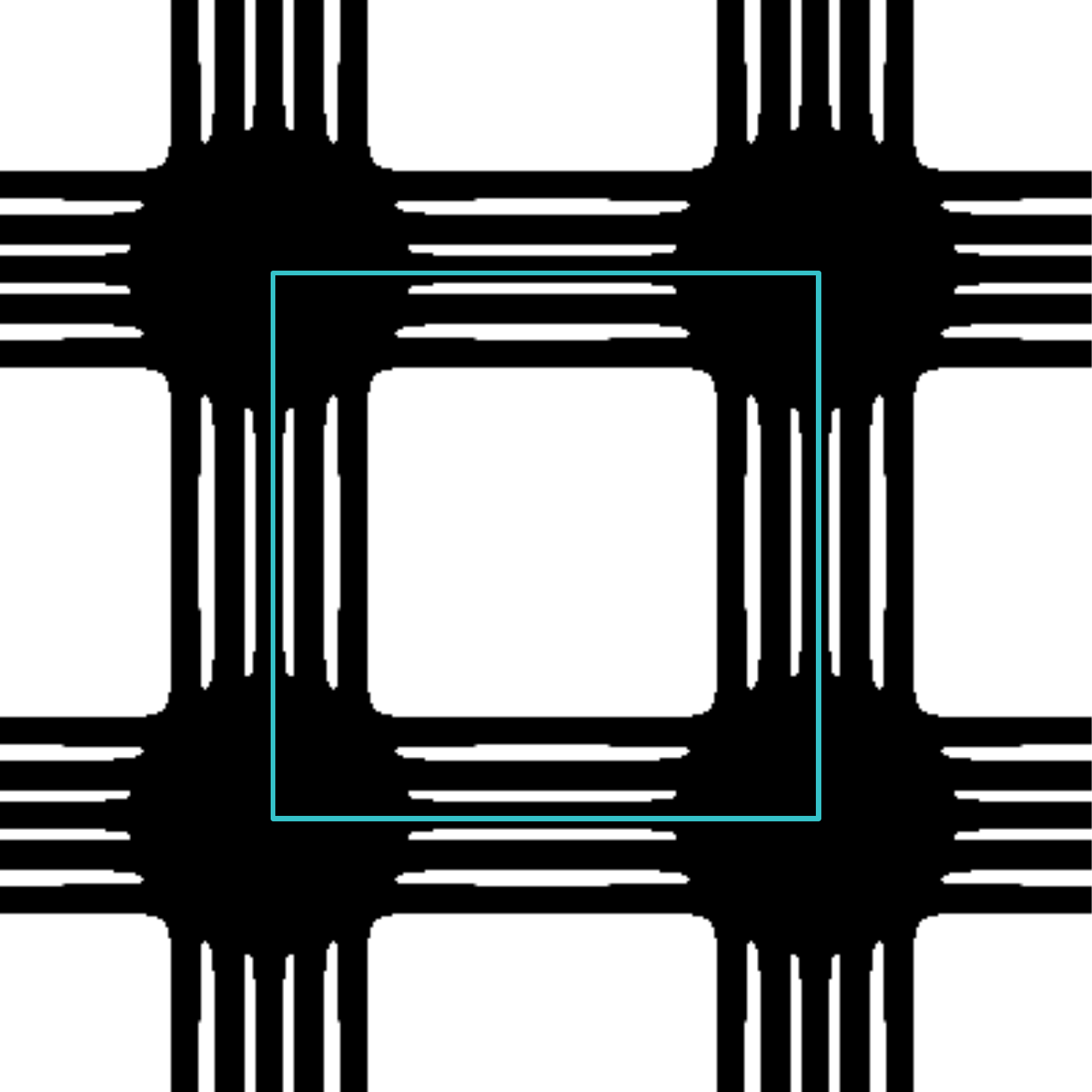}} \\
\caption{Resulting structures shown in a domain of size $2 \times 2$ with the unit-cell highlighted, for $\chi=0.0$, $f=0.5$, and a length-scale of $0.05$.}
\label{Fig:Example.4}
\end{figure}

An interesting observation is made from comparing the optimized designs using \textit{Mapped SG} and \textit{Homog. SG}  obtained for $\chi=0$, shown in Figure~\ref{Fig:Example.4}(a) and (b). Here, the \textit{Mapped SG} got stuck in the slightly suboptimal single-length-scale design dictated by the starting guess, whereas the \textit{Homog. SG} design resulted in a split bar structure, somewhat mimicking the special class of extremal composites proposed by~\citet{Bib:SigmundExtremal2000}. Actually, the ``true optimal'' single-scale microstructure will probably be one where the width and number of the split bars are exactly determined by the imposed minimum length-scale. This can be mimicked by modifying the \textit{Mapped SG} starting guess to accommodate for these extremal composites. In practice, this is done by splitting thick members of the mapped design into members that are close to the minimum feature size as seen in Figure~\ref{Fig:Example.4}(c). As expected, the relative performance of the improved mapping, which is $1.10$ is lower than the value of $1.12$ for the \textit{Mapped Rank-3}. The corresponding optimized design, shown in Figure~\ref{Fig:Example.4}(d) resembles the extremal Sigmund composite, and its normalized performance of $1.03$ outperforms the microstructure obtained using the \textit{Homog. SG}, which has a relative performance of $1.04$. Nevertheless, we abandon this more advanced starting guess in the following, in favor of producing simpler and easier to manufacture designs.

\subsection{Example 2: Large Filter}
This example uses the same load case and volume fraction as Example 1, but imposes a larger minimal length-scale of $0.15$, corresponding to 15\% of the square domain. Hence, the microstructures optimized using \textit{Random SG's} and \textit{Homog. SG's} are forced to find structures with thicker features, closer to the feature sizes of the \textit{Mapped SG}.

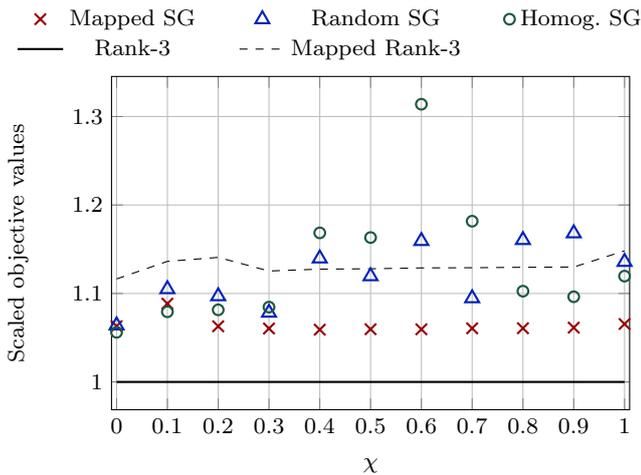
\begin{figure}
\centering
\begin{tikzpicture}[scale=1]
\footnotesize
\begin{axis}[
    		    grid=major,
                width=\linewidth, 
                height=6cm,
                scaled x ticks = false,
                /pgf/number format/.cd,
                set thousands separator={ },
                xlabel = {$\chi$},
                ylabel = {Scaled objective values},
                xmin=-0.01,xmax=1.01,
                legend entries={
                    Mapped SG,
                    Random SG,
                    Homog. SG,
                    Rank-3,
                    Mapped Rank-3
                },
                legend style={
                    draw=none,
                    legend columns=3,
                    at={(1.05,1.02)},
                    anchor=south east,
                    /tikz/every even column/.append style={column sep=0.5cm}
                },
                xtick = {0,.1,.2,.3,0.4,.5,0.6,.7,0.8,.9,1}
    ] 
        
    
    \addplot[mark=x,mark size=3,NewRed,only marks,thick] table[x index=0, y index=3] {ex2.txt};
    
    \addplot[mark=triangle,mark size=3,NewBlue,only marks,thick] table[x index=0, y index=4] {ex2.txt};
    
    \addplot[mark=o,mark size=2,NewTeal,only marks,thick] table[x index=0, y index=5] {ex2.txt};
    
    \addplot[sharp plot,black,thick] table[x index=0, y index=1] {ex2.txt};
    
    \addplot[sharp plot,dashed,black] table[x index=0, y index=2] {ex2.txt};

\end{axis}
\end{tikzpicture}
\caption{$\mathcal{C}$ normalized using the rank-3 energy bound, for different values of $\chi$, $f=0.5$, and a length-scale of $0.15$.}
\label{Fig:Example.5}
\end{figure}

The relative objective function values are shown in Figure~\ref{Fig:Example.5}. It is clear that designs optimized using \textit{Mapped SG} now outperform the microstructures using \textit{Random SG's} and \textit{Homog. SG's} in terms of objective function value. Some optimized designs using \textit{Random SG's} and \textit{Homog. SG's} are even outperformed by the \textit{Mapped Rank-3}, which is obtained using purely geometrical means. Finally, it can be seen from Figures~\ref{Fig:Example.2} and~\ref{Fig:Example.5} that both the mapped structures and subsequently optimized microstructures have a very stable relative objective function value across the load cases, compared to the other starting guesses.

\begin{figure}[h!]
\centering
\subfloat[Mapped SG.]{\includegraphics[width=0.45\linewidth]{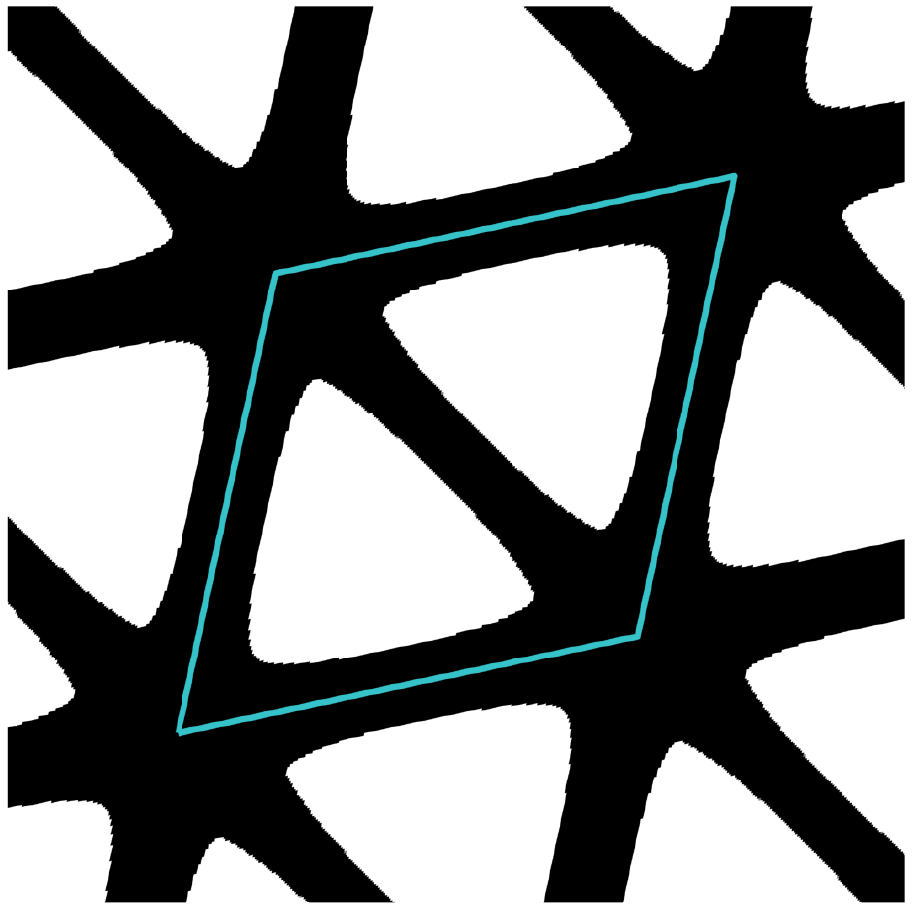}} \quad
\subfloat[Homog. SG.]{\includegraphics[width=0.45\linewidth]{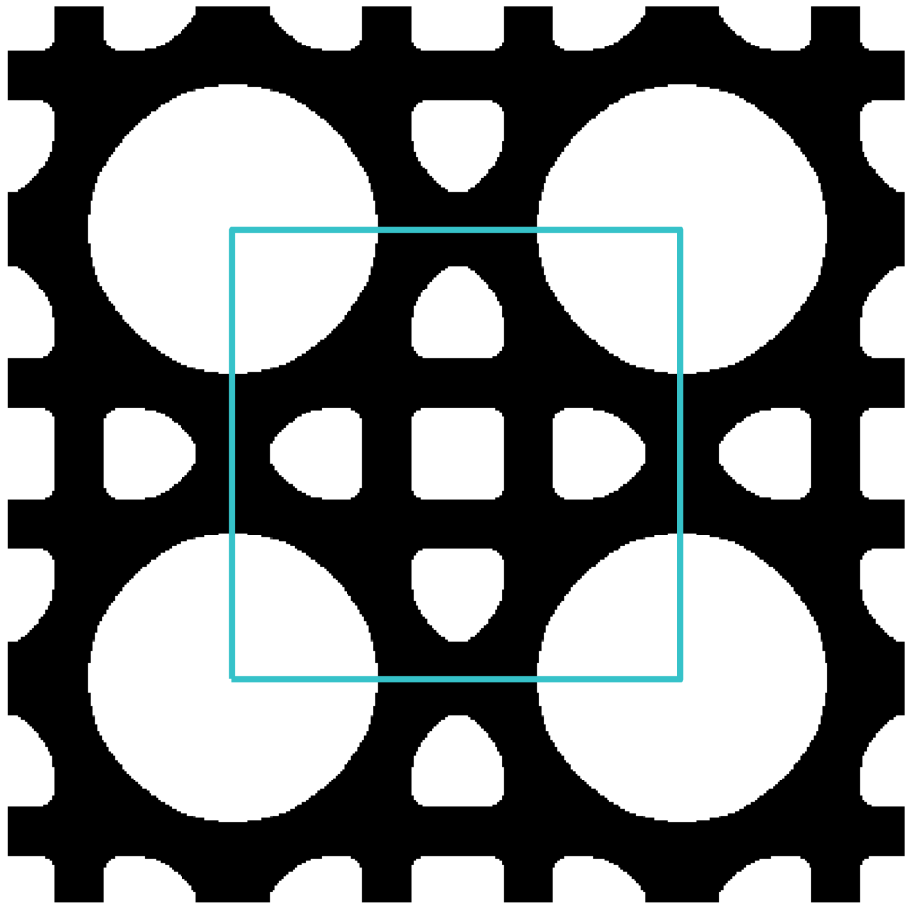}} \\
\caption{Resulting structures shown in a domain of size $2 \times 2$ with the unit-cell highlighted, for $\chi=0.3$, $f=0.5$, and a length-scale of $0.15$.}
\label{Fig:Example.6}
\end{figure}

When comparing the resulting structures for $\chi =0.3$, shown in Figure~\ref{Fig:Example.6}, it can be seen that the structure obtained by the \textit{Mapped SG} is still very simple compared to the one obtained by the \textit{Homog. SG}, as the latter contains smaller holes.

\subsection{Example 3: Low volume fraction}
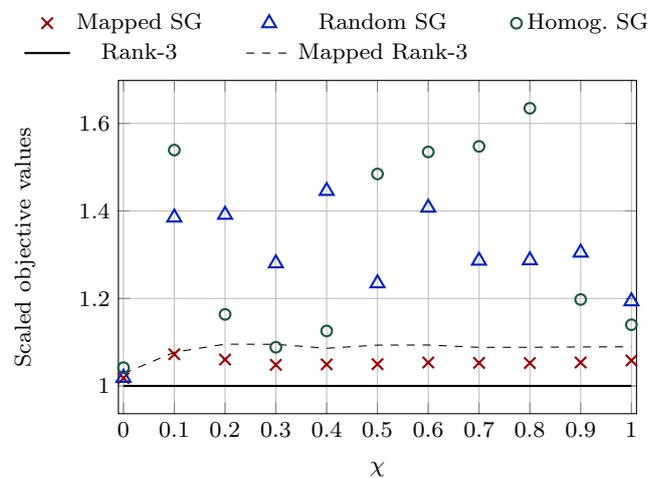
\begin{figure}
\centering
\begin{tikzpicture}[scale=1]
\footnotesize
\begin{axis}[
    		    grid=major,
                width=\linewidth, 
                height=6cm,
                scaled x ticks = false,
                /pgf/number format/.cd,
                set thousands separator={ },
                xlabel = {$\chi$},
                ylabel = {Scaled objective values},
                xmin=-0.01,xmax=1.01,
                legend entries={
                    Mapped SG,
                    Random SG,
                    Homog. SG,
                    Rank-3,
                    Mapped Rank-3
                },
                legend style={
                    draw=none,
                    legend columns=3,
                    at={(1.05,1.02)},
                    anchor=south east,
                    /tikz/every even column/.append style={column sep=0.5cm}
                },
                xtick = {0,.1,.2,.3,0.4,.5,0.6,.7,0.8,.9,1}
    ] 
        
    
    \addplot[mark=x,mark size=3,NewRed,only marks,thick] table[x index=0, y index=3] {ex3.txt};
    
    \addplot[mark=triangle,mark size=3,NewBlue,only marks,thick] table[x index=0, y index=4] {ex3.txt};
    
    \addplot[mark=o,mark size=2,NewTeal,only marks,thick] table[x index=0, y index=5] {ex3.txt};
    
    \addplot[sharp plot,black,thick] table[x index=0, y index=1] {ex3.txt};
    
    \addplot[sharp plot,dashed,black] table[x index=0, y index=2] {ex3.txt};

\end{axis}
\end{tikzpicture}
\caption{$\mathcal{C}$ normalized using the rank-3 energy bound, for different values of $\chi$, $f=0.2$, and a length-scale of $0.05$.}
\label{Fig:Example.7}
\end{figure}

The same load cases are used as for the previous two examples; however, in this example a lower volume fraction of $f=0.2$ is used in combination with a minimum feature size of $0.05$. The resulting performance values are shown in Figure~\ref{Fig:Example.7}. It can be seen that the results using the \textit{Mapped SG} clearly outperform the other starting guesses, except for the case with $\chi=0$, where all methods perform equally well.

\begin{figure}[h!]
\centering
\subfloat[Mapped Rank-3.]{\includegraphics[width=0.45\linewidth]{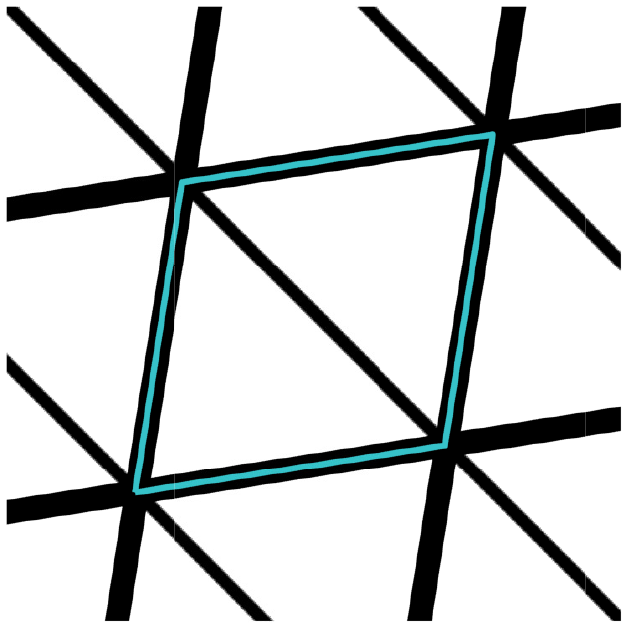}} \quad
\subfloat[Mapped SG.]{\includegraphics[width=0.45\linewidth]{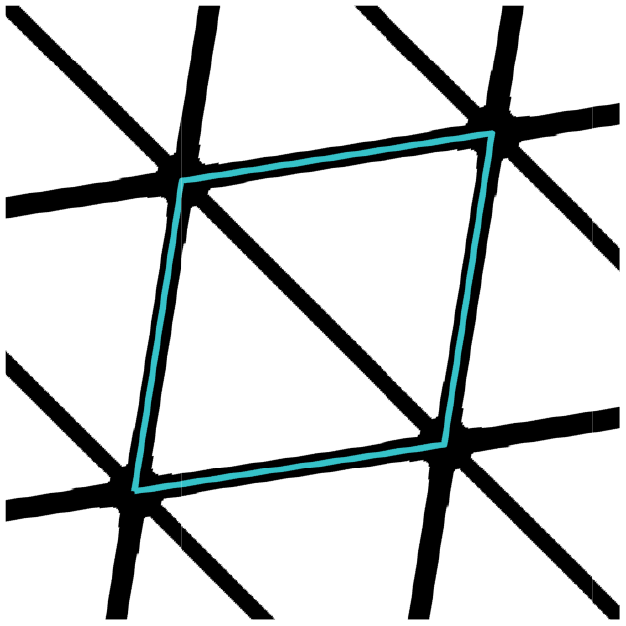}} \\
\caption{Resulting structures shown in a domain of size $2 \times 2$ with the unit-cell highlighted, for $\chi=0.1$, $f=0.2$, and a length-scale of $0.05$.}
\label{Fig:Example.8}
\end{figure}

\begin{figure}[h!]
\centering
\subfloat[Mapped SG.]{\includegraphics[width=0.45\linewidth]{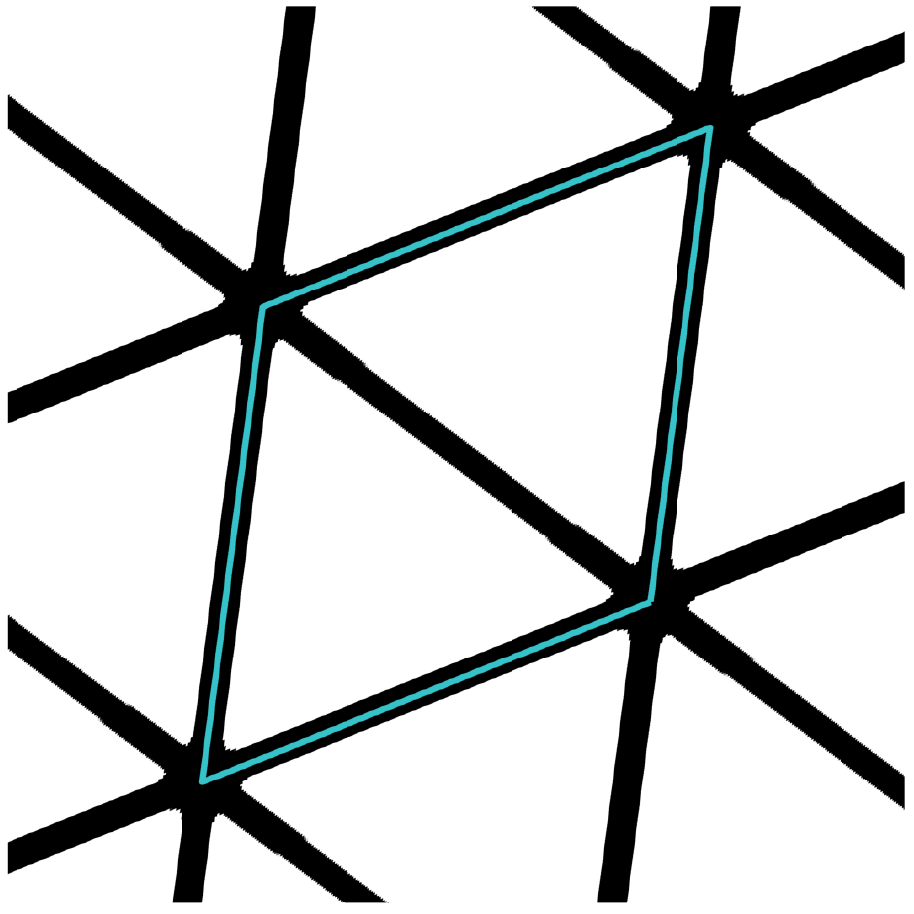}} \quad
\subfloat[Homog. SG.]{\includegraphics[width=0.45\linewidth]{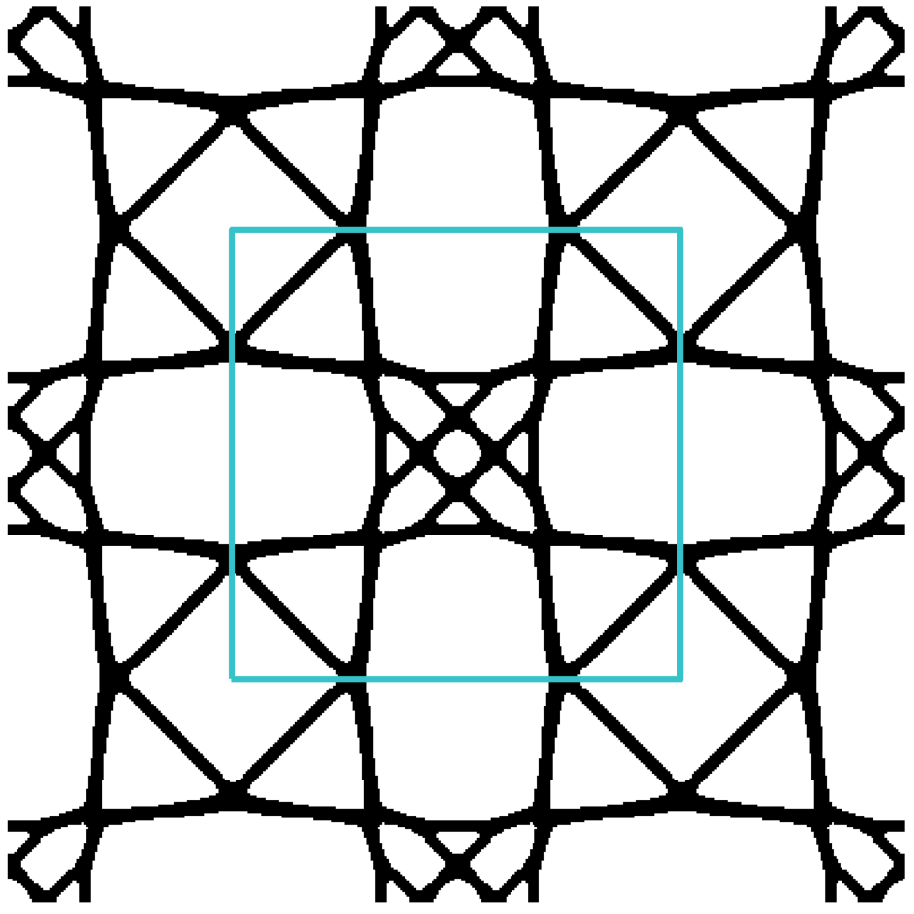}} \\
\caption{Resulting structures shown in a domain of size $2 \times 2$ with the unit-cell highlighted, for $\chi=0.8$, $f=0.2$, and a length-scale of $0.15$.}
\label{Fig:Example.9}
\end{figure}

When inspecting the resulting structures shown for $\chi=0.1$ in Figure~\ref{Fig:Example.8} and for $\chi=0.8$
in Figure~\ref{Fig:Example.9}, it can again be seen that the designs based on the \textit{Mapped SG} are simple triangle structures. However, it can also be seen that the diagonal bar crossing the center of the unit-cell in Figure~\ref{Fig:Example.8}(b) has exactly the imposed minimum length-scale as feature width. Interestingly, this bar is thicker than the bar following directly from \textit{Mapped Rank-3}, as shown in Figure~\ref{Fig:Example.8}(a). Hence, the imposed feature size limits the optimality of the optimized microstructure. This effect can also be seen in Figure~\ref{Fig:Example.7} where the relative compliance value is slightly higher for $\chi = 0.1$ than for  all other values of $\chi$. A similar effect can be observed for $\chi=0.1$ in Figure~\ref{Fig:Example.5}.

\subsection{Example 4: Rotating uni-axial stresses}
\begin{figure}
\centering
\includegraphics[width=.95\linewidth]{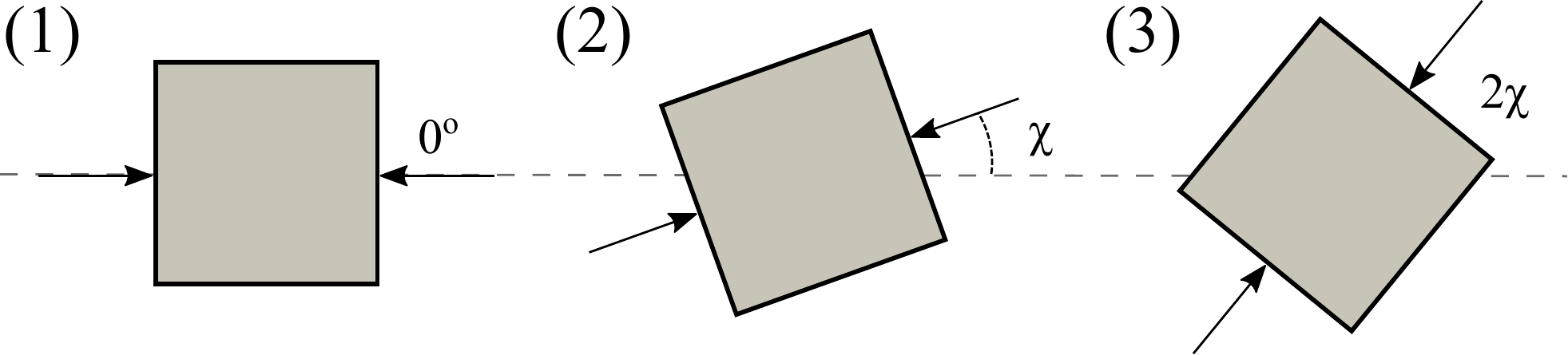}
\caption{Illustration of three uni-axial loads, with angles 0, $\chi$, and $2 \chi$.}
\label{Fig:Example.10}
\end{figure}

For this final example we introduce a new loading case with three uni-axial stresses, independently applied in different directions as illustrated in Figure~\ref{Fig:Example.10}. The angle parameter $\chi$ is increased in steps of 5 degrees from 0 to 60 degrees, resulting in a load case varying from a single load to three uni-axial loads with $60^{\circ}$ symmetry. The example is performed with a volume fraction of $f=0.25$, using a minimum feature length of $0.05$.

\begin{figure}
\centering
\begin{tikzpicture}[scale=1]
\footnotesize
\begin{axis}[
    		    grid=major,
                width=\linewidth, 
                height=6cm,
                scaled x ticks = false,
                /pgf/number format/.cd,
                set thousands separator={ },
                xlabel = {$\chi$ (in degrees)},
                ylabel = {Scaled objective values},
                xmin=-0.01,xmax=60.01,
                legend entries={
                    Mapped SG,
                    Random SG,
                    Homog. SG,
                    Rank-3,
                    Mapped Rank-3
                },
                legend style={
                    draw=none,
                    legend columns=3,
                    at={(1.05,1.02)},
                    anchor=south east,
                    /tikz/every even column/.append style={column sep=0.5cm}
                },
                xtick = {0,10,20,30,40,50,60}
    ] 
        
    
    \addplot[mark=x,mark size=3,NewRed,only marks,thick] table[x index=0, y index=3] {ex4.txt};
    
    \addplot[mark=triangle,mark size=3,NewBlue,only marks,thick] table[x index=0, y index=4] {ex4.txt};
    
    \addplot[mark=o,mark size=2,NewTeal,only marks,thick] table[x index=0, y index=5] {ex4.txt};
    
    \addplot[sharp plot,black,thick] table[x index=0, y index=1] {ex4.txt};
    
    \addplot[sharp plot,dashed,black] table[x index=0, y index=2] {ex4.txt};

\end{axis}
\end{tikzpicture}
\caption{$\mathcal{C}$ normalized using the rank-3 energy bound, for different values of $\chi$ corresponding to example 4, $f=0.25$, and a length-scale of $0.05$.}
\label{fig:ex4_sweep}
\end{figure}
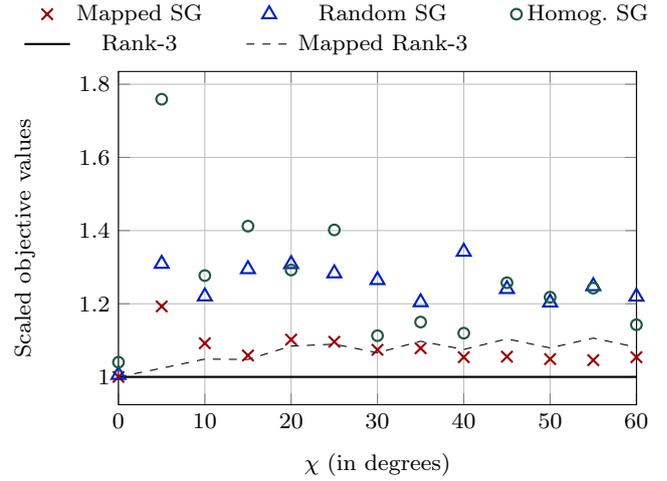

\begin{figure}[h!]
\centering
\subfloat[Mapped Rank-3.]{\includegraphics[width=0.45\linewidth]{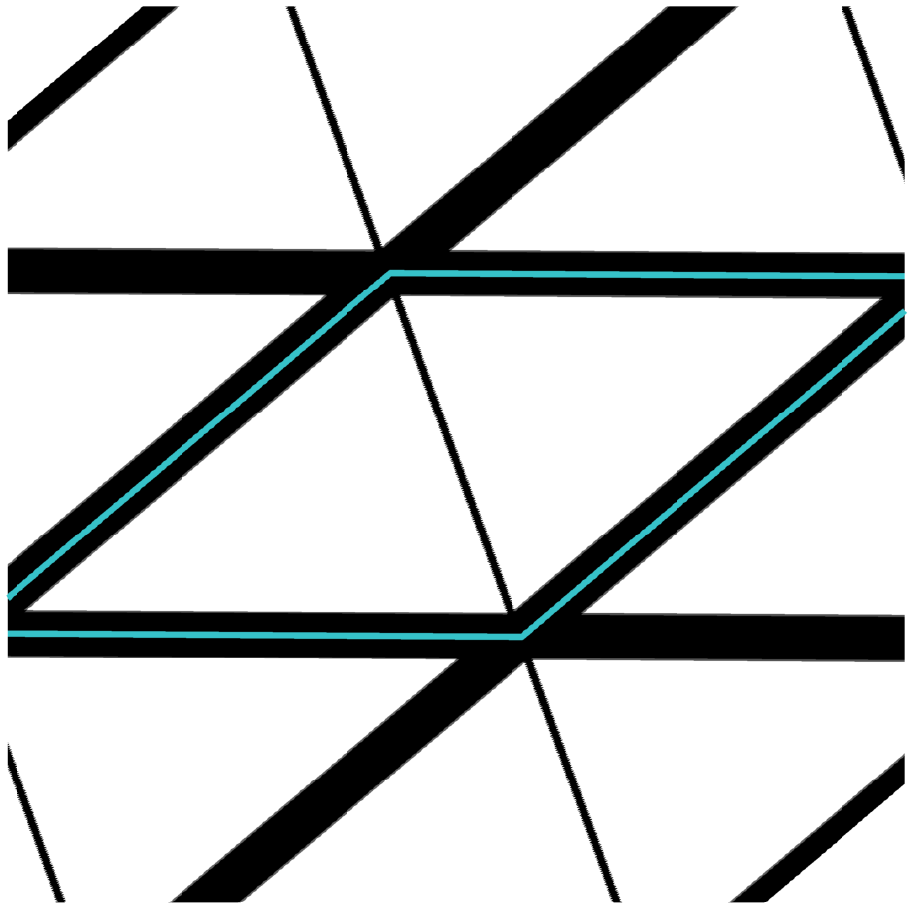}} \quad
\subfloat[Mapped SG.]{\includegraphics[width=0.45\linewidth]{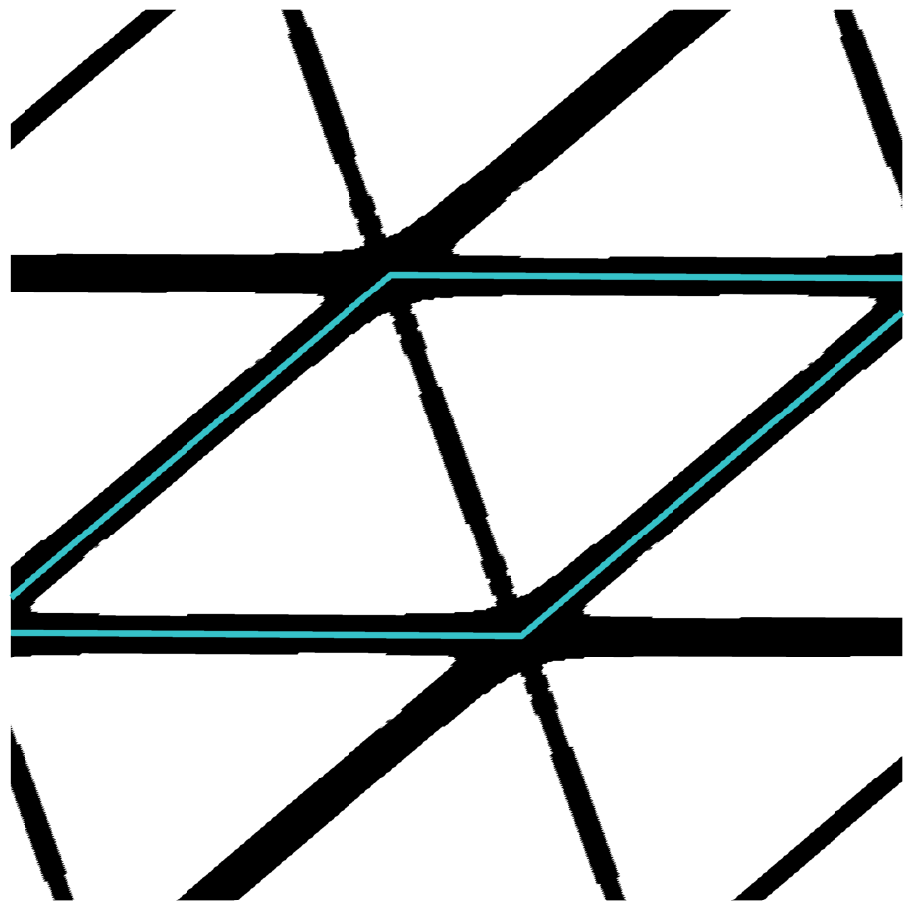}} \\
\subfloat[Random SG.]{\includegraphics[width=0.45\linewidth]{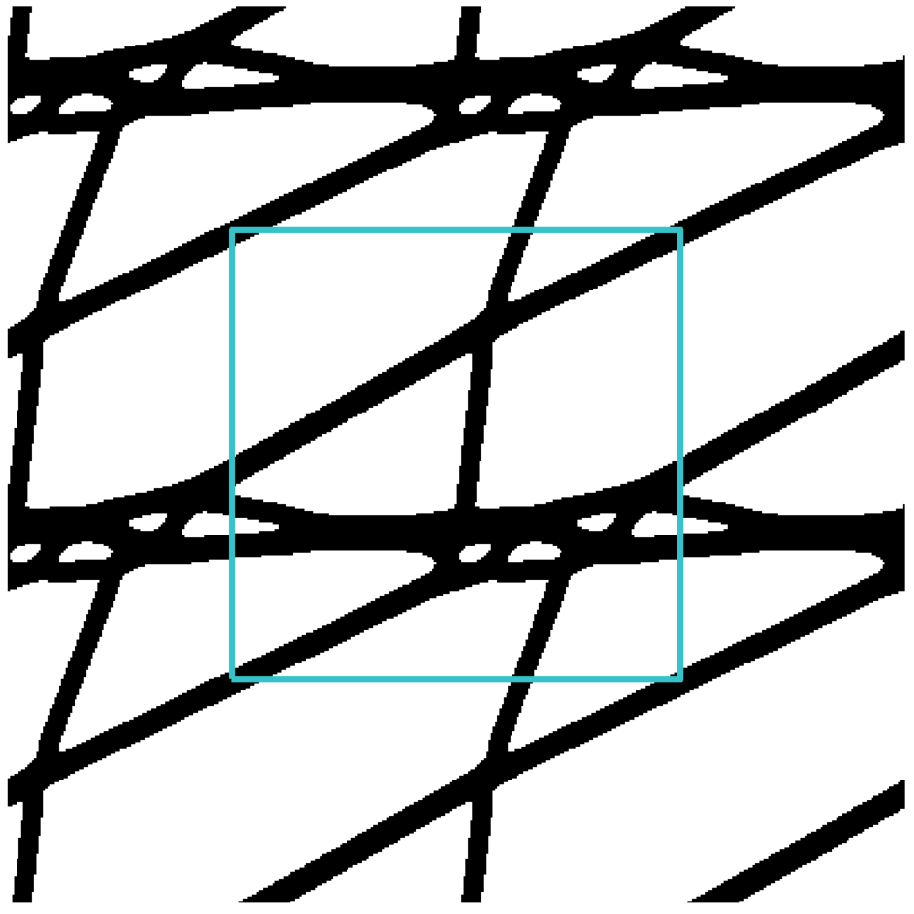}}  \quad
\subfloat[Homog. SG.]{\includegraphics[width=0.45\linewidth]{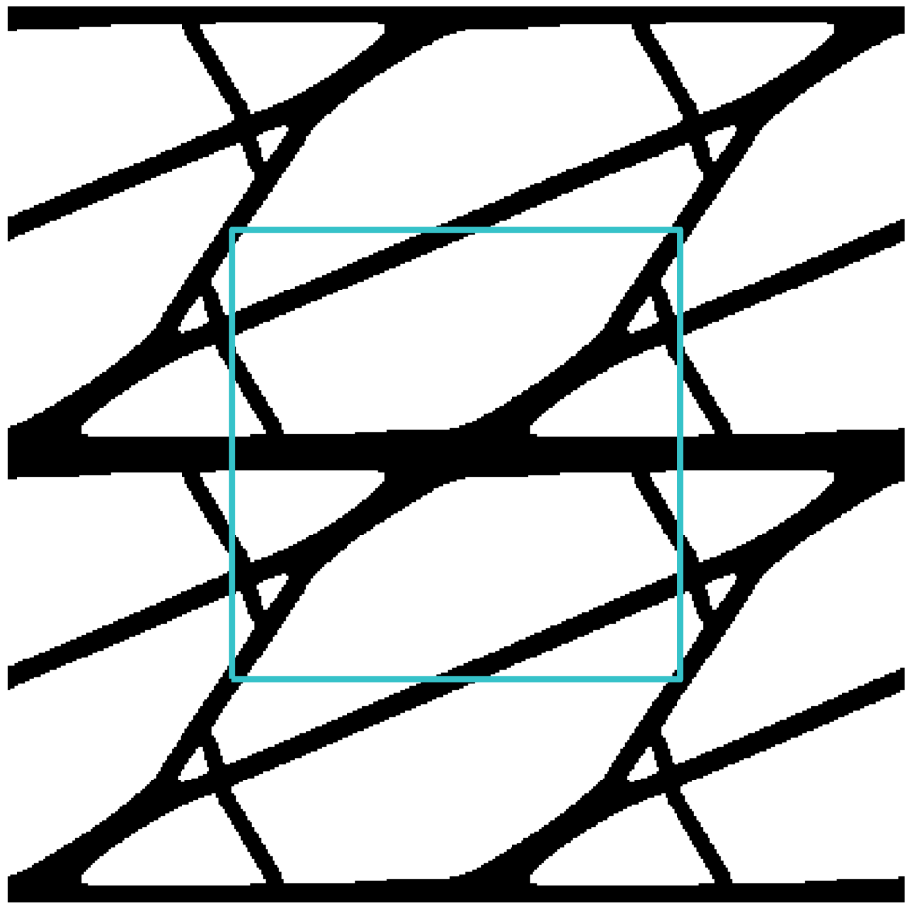}} \\
\caption{Resulting structures shown in a domain of size $2 \times 2$ with the unit-cell highlighted, for $\chi=20^{\circ}$, $f=0.25$, and a length-scale of $0.05$.}
\label{Fig:Example.12}
\end{figure}

The resulting normalized objective function values are shown in \ref{fig:ex4_sweep}, where it can be seen that using the \textit{Mapped SG} results in much better objective function values for all cases. Another important observation is that the \textit{Mapped Rank-3} outperforms the optimized structure for values of $\chi = 5^{\circ},10^{\circ},15^{\circ},20^{\circ},25^{\circ},30^{\circ}$. This is, as discussed before, due to the minimum length-scale not being enforced in the reconstruction of the rank-3 laminate, as is shown in Figure~\ref{Fig:Example.12}. After the topology optimization procedure the concerned members have either the imposed length-scale or are removed completely.

\section{Discussion and conclusions}
\label{Sec:Discussion}
When comparing the various starting guesses as applied here, it is important to note that two different unit-cell parameterizations are used. Due to the parallelogram element shapes, features will have different jaggedness in the solid-void interface depending on orientation. To avoid negative effects caused by this difference the optimization problems have been resolved using a sufficiently fine mesh such that the minimum feature size is at least 10 finite elements wide in all examples. Furthermore, the skewness of the parallelogram used in the \textit{Mapped SG} makes it impossible to recreate the same structure in a square domain, as they are repeated differently through space. The choice of using a square unit-cell for the other two starting guesses was made since this resulted in the best performing microstructures.

The objective values obtained using the \textit{Mapped SG} display a constant excellent performance, where the results are always around 5-8$\%$ from the energy bound. Only when the imposed feature size is larger than a member needs to be, this closeness to theoretical optimality is violated. Furthermore, the optimized microstructures using \textit{Mapped SG} are much simpler than their counterparts optimized using \textit{Random SG} or \textit{Homog. SG}. This simplicity in the microstructures possibly allows for a more robust behavior towards manufacturing uncertainties, and for a simpler manufacturing process in general.

Only in Example 1, it was shown that the use of the \textit{Homog. SG} could lead to a slightly better performance. The reason for this, is that this starting guess does not suppress the occurrence of thin features, as happens using \textit{Mapped SG}. Hence, the optimized microstructures based on \textit{Homog. SG} seem to converge better towards the type of extremal composites proposed by~\citet{Bib:SigmundExtremal2000}. However, we also demonstrated that the \textit{Mapped SG} could be improved when the minimum allowed feature size allows for a representation close to Sigmund's class of extremal materials. Nevertheless, we did not pursue this approach further in this work, since we believe that the simplicity of the microstructures using the \textit{Mapped SG} are an important feature. Furthermore, it is noteworthy that for Examples 2, 3 and 4, large differences are observed between performance of \textit{Mapped SG} and the other two starting guesses. Hence, it really pays off to use a ``smart" starting guess as suggested here. Especially since the cost of obtaining the approximated rank-3 microstructure on the single scale, is negligible compared to the cost of performing inverse homogenization.

The approximated rank-3 laminates, directly analyzed and without optimization (\textit{Mapped Rank-3}), already perform within $5$-$15\%$ of the energy bound posed by the rank-3 laminate. This promising result justifies the recent interest into projection methods that post-process homogenization-based topologies on a single scale~\citep{Bib:PantzTrabelsi,Bib:GroenSigmund2017,BiB:GDondersAllairePantz2018}. This result also indicates that similar performance close to theoretical optima can be obtained when multiple load cases are considered. Furthermore, it is notable that the main difference between the \textit{Mapped Rank-3} and the microstructures optimized from the \textit{Mapped SG} are rounded corners at the intersection of connecting bars. This indicates that some more or less heuristic post-processing scheme may be postulated, that is applied directly on the \textit{Mapped Rank-3}, potentially eliminating the need for inverse homogenization entirely.

\begin{acknowledgements}
The authors acknowledge the support of the Villum Fonden through the Villum investigator project InnoTop. The authors would also like to thank Krister Svanberg for providing the MATLAB MMA code.
\end{acknowledgements}

\appendix
\section*{Appendix: Reconstructing a rank-3 laminate from moments}
\label{Sec:Appendix.recon}
In this section we discuss the method proposed by \citet{Bib:Lipton1994MomentRecon} to reconstruct the relative layer widths $\mu_{n}$ and orientations $\theta_{n}$ of a rank-3 laminate from the optimal moments. Our discussion is similar to the practical implementation given by \citet{Bib:DiazLiptonSoto94}, and included for completeness.

\begin{figure}[h!]
\centering
\includegraphics[width=0.5\textwidth]{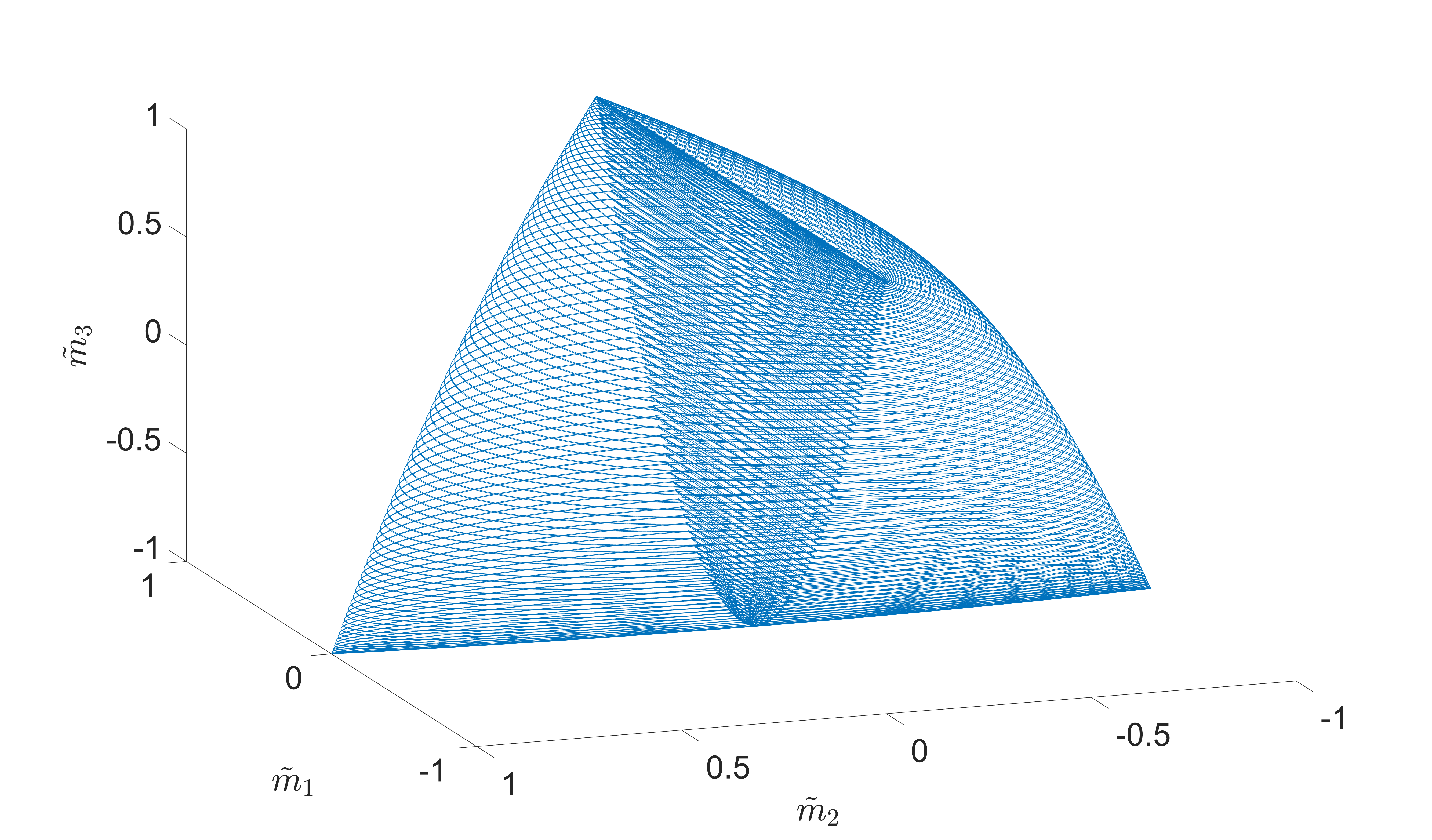}
\caption{Convex set $\tilde{\mathcal{M}}$.} 
\label{Fig:recon.1}
\end{figure}

We can reduce the set of optimal moments from four to three, by rotating the set of moments $(m_{1},m_{2},m_{3},m_{4})$ to $(\tilde{m}_{1},\tilde{m}_{2},\tilde{m}_{3},0)$ using a change of reference frame and rotation angle $\gamma$ such that $\tilde{\theta}_{n} = \theta_{n}+\gamma$ and,
\begin{equation} \label{Eq:recon.1}
    \begin{aligned}
    \tilde{m}_{1} &= \sum_{n=1}^{N}p_{n} \text{cos}(2\tilde{\theta}_{n}), && \tilde{m}_{2} = \sum_{n=1}^{N}p_{n} \text{sin}(2\tilde{\theta}_{n}), \\
    \tilde{m}_{3} &= \sum_{n=1}^{N}p_{n} \text{cos}(4\tilde{\theta}_{n}), && \tilde{m}_{4} = \sum_{n=1}^{N}p_{n} \text{sin}(4\tilde{\theta}_{n}). \\
	\end{aligned}
\end{equation}
By using the rotated reference frame for the specification of the layer tangents $\boldsymbol{t}_{n}$, used in Equation~\ref{Eq:Moment.5}, the following relations can be found,
\begin{equation} \label{Eq:recon.5}
    \begin{aligned}
    \tilde{m}_{1} &= m_{1}\text{cos}(2\gamma) - m_{2}\text{sin}(2\gamma),\\
	\tilde{m}_{2} &= m_{1}\text{sin}(2\gamma) + m_{2}\text{cos}(2\gamma), \\
	\tilde{m}_{3} &= m_{3}\text{cos}(4\gamma) - m_{4}\text{sin}(4\gamma), \\
	\tilde{m}_{4} &= m_{3}\text{sin}(4\gamma) + m_{4}\text{cos}(4\gamma). \\
	\end{aligned}
\end{equation}
Hence, we can find $\gamma$ that ensures $\tilde{m}_{4}=0$ using,
\begin{equation} \label{Eq:recon.4}
\gamma = \frac{1}{4} \text{arctan}(\frac{-m_{4}}{m_{3}}).
\end{equation}
From Equation \ref{Eq:recon.4} it can be seen that $\gamma$ is periodic every $\pi/4$. This means that there are at least four rotated sets $\tilde{\boldsymbol{m}}$ to describe the microstructure. Furthermore, the feasible rotated set of moments $\tilde{\mathcal{M}}$ is bounded by the same constraints as in Equation~\ref{Eq:recon.6}. 
\begin{equation}\label{Eq:recon.6}
\tilde{\mathcal{M}} = \tilde{\boldsymbol{m}} \in \mathbb{R}^{3},   s.t. \begin{cases}
\tilde{m}_{1}^{2} + \tilde{m}_{2}^{2} \leq 1,  \\
-1 \leq \tilde{m}_{3} \leq 1, \\
\frac{2\tilde{m}_{1}^{2}}{1+\tilde{m}_{3}}+\frac{2\tilde{m}_{2}^{2}}{1-\tilde{m}_{3}} \leq 1.
\end{cases}
\end{equation}
Feasible set $\tilde{\mathcal{M}}$ is a convex set as can be seen in Figure~\ref{Fig:recon.1}. 

The boundary of this convex set $\partial\tilde{\mathcal{M}}$ satisfies,
\begin{equation}\label{Eq:recon.7}
\frac{2\tilde{m}_{1}^{2}}{1+\tilde{m}_{3}}+\frac{2\tilde{m}_{2}^{2}}{1-\tilde{m}_{3}} = 1,
\end{equation}
while the four corner points also satisfy,
\begin{equation}\label{Eq:recon.8}
\tilde{m}_{1}^{2} + \tilde{m}_{2}^{2} = 1.
\end{equation}
It can easily be verified using Equations~\ref{Eq:recon.1} that the four corner points $\tilde{\boldsymbol{m}} = \left\{1,0,1\right\}$, $\left\{-1,0,1\right\}$, $\left\{0,1,-1\right\}$ and $\left\{0,-1,-1\right\}$ correspond to rank-1 laminates, with the corresponding layer directions $\tilde{\theta}_{1}=0$, $\pi/2$, $\pi/4$ and $-\pi/4$ respectively. Hence, if both Equations~\ref{Eq:recon.7} and \ref{Eq:recon.8} are satisfied the microstructure is a rank-1 microstructure. Depending on the choice of $\gamma$ the unique layer orientation $\theta_{1}$ can be obtained; furthermore, corresponding $p_{1}=1$.

Since $\tilde{\mathcal{M}}$ is a convex set, each point $\tilde{\boldsymbol{m}}$ can be described as a combination of a corner point $\tilde{\boldsymbol{a}}$ and a point $\tilde{\boldsymbol{b}}$ on $\partial\tilde{\mathcal{M}}$,
\begin{equation}\label{Eq:recon.9}
\tilde{\boldsymbol{m}} = \alpha\tilde{\boldsymbol{a}} +(1-\alpha)\tilde{\boldsymbol{b}}.
\end{equation}
Since $\tilde{\boldsymbol{a}}$ corresponds to a rank-1 laminate, point $\tilde{\boldsymbol{b}}$ on boundary $\partial\tilde{\mathcal{M}}$ has to correspond to a rank-2 laminate. Hence,  
\begin{equation}\label{Eq:recon.10}
\frac{2\tilde{b}_{1}^{2}}{1+\tilde{b}_{3}}+\frac{2\tilde{b}_{2}^{2}}{1-\tilde{b}_{3}} = 1.
\end{equation}
Furthermore, we have
\begin{equation}\label{Eq:recon.11}
\alpha = \frac{\tilde{m}_{1}-\tilde{b}_{1}}{\tilde{a}_{1}-\tilde{b}_{1}} = \frac{\tilde{m}_{2}-\tilde{b}_{2}}{\tilde{a}_{2}-\tilde{b}_{2}} =\frac{\tilde{m}_{3}-\tilde{b}_{3}}{\tilde{a}_{3}-\tilde{b}_{3}}.
\end{equation}
If we take one of the corner points, e.g. $\tilde{\boldsymbol{a}} = \left\{1,0,1\right\}$, we can solve for $\tilde{\boldsymbol{b}}$ and $\alpha$, using the Equations above. We know that the rank-2 laminate can be described using two relative layer contributions $p_{1}^{\tilde{b}}$ and $p_{2}^{\tilde{b}}$, and two angles $\theta_{1}^{\tilde{b}}$ and $\theta_{2}^{\tilde{b}}$, such that,
\begin{equation}\label{Eq:recon.12}
\begin{aligned}
p_{1}^{\tilde{b}} & + p_{2}^{\tilde{b}} = 1, \\
\tilde{b}_{1} & = p_{1}^{\tilde{b}} \text{cos}(2\theta_{1}^{\tilde{b}}) + p_{2}^{\tilde{b}} \text{cos}(2\theta_{2}^{\tilde{b}}),\\
\tilde{b}_{2} & = p_{1}^{\tilde{b}} \text{sin}(2\theta_{1}^{\tilde{b}}) + p_{2}^{\tilde{b}} \text{sin}(2\theta_{2}^{\tilde{b}}), \\
\tilde{b}_{3} & = p_{1}^{\tilde{b}} \text{cos}(4\theta_{1}^{\tilde{b}}) + p_{2}^{\tilde{b}} \text{cos}(4\theta_{2}^{\tilde{b}}). 
\end{aligned}
\end{equation}
This is system of four equations can be solved for the four unknowns. To do so, one can describe $\tilde{\boldsymbol{b}}$ in terms of two angles, $0\leq t \leq 2\pi$ and $0\leq \beta \leq \pi/2$~\citep{Bib:Lipton1994MomentRecon}.
\begin{equation}\label{Eq:recon.13}
\left\{\tilde{b}_{1},\tilde{b}_{2},\tilde{b}_{3}\right\} = \left\{\text{cos}(\beta)\text{cos}(t),\text{sin}(\beta)\text{sin}(t),\text{cos}(2\beta) \right\}.
\end{equation}
The corresponding solution for the rank-2 laminate can then be written as,
\begin{equation}\label{Eq:recon.14}
\begin{aligned}
s & = \text{cos}^{2}(2\beta)-2\text{cos}(2\beta)\text{cos}(2t)+1, \\
\delta & = \text{arctan}\Big(\frac{\sqrt{s}}{1-\text{cos}^{2}(2\beta)} \Big), \\
p_{1}^{\tilde{b}} & = \frac{1}{2}\Big( 1 + \frac{2\text{cos}(2\beta)\text{sin}(2t)}{\sqrt{s}} \Big), \\
p_{2}^{\tilde{b}} & = 1 - p_{1}^{\tilde{b}}, \\
\theta_{1}^{\tilde{b}} & = \text{arctan}\Big( \frac{p_{2}^{\tilde{b}}\text{sin}(2\delta)-\text{sin}(\beta)\text{sin}(t)}{-(p_{1}^{\tilde{b}}+p_{2}^{\tilde{b}}\text{cos}(2\delta)+\text{cos}(\beta)\text{cos}(t))} \Big), \\
\theta_{2}^{\tilde{b}} & = \theta_{1}^{\tilde{b}}  + \delta.
\end{aligned}
\end{equation}
The corresponding rank-3 laminate in global frame of reference can thus be written as,
\begin{equation}\label{Eq:recon.15}
\begin{aligned}
p_{1} & = \alpha, 						&& \theta_{1} = -\gamma,& \\
p_{2} & = (1-\alpha)p_{1}^{\tilde{b}}, && \theta_{2} = \theta_{1}^{\tilde{b}}-\gamma,& \\
p_{3} & = (1-\alpha)p_{2}^{\tilde{b}}, && \theta_{3} = \theta_{2}^{\tilde{b}}-\gamma.& 
\end{aligned}
\end{equation}
Finally, the relative widths used at each layer can be obtained using,
\begin{equation}\label{Eq:recon.16}
\begin{aligned}
\mu_{1} &= p_{1} f, \\
\mu_{2} &= \frac{p_{2} f}{1-\mu_{1}},\\
\mu_{3} &= \frac{p_{3} f}{1-(\mu_{1}+\mu_{2}-\mu_{1}\mu_{2})}.
\end{aligned}
\end{equation}


\bibliography{biblio}
\bibliographystyle{elsarticle-harv}

\end{document}